\documentclass[usenatbib,usegraphicx]{mn2e}
\usepackage{aas_macros}
\usepackage{natbib}
\usepackage{amsmath,amssymb,amsfonts,bm}
\usepackage{subfigure}
\usepackage{longtable}
\usepackage{graphicx}
\usepackage{graphics}
\usepackage{keyval}
\usepackage{trig}
\usepackage{float}
\usepackage{calc}
\usepackage{url}
\usepackage{hyperref}
\def\beq{\begin{equation}}
\def\eeq{\end{equation}}
\def\bey{\begin{eqnarray}}
\def\eey{\end{eqnarray}}
\def\Myr{\, {\rm Myr} }

\def\pc{\, {\rm pc} }

\def\kpc{\, {\rm kpc} }
\def\msun{M_\odot}
\def\sun{\odot}
\def\Msun{M_\odot}

\def\kms{\, {\rm km \, s}^{-1} }

\def\dt{{\bf d}t}
\def\grad{{\bf \nabla}}

\def\a0{$a_0$}

\def\tcm{{\rm T}_{\rm cross}^{\rm MD}}
\def\tcn{{\rm T}_{\rm cross}^{\rm Newt}}
\def\vir{\rm 2K/|W|}
\def\ge{{\bf g}_{\rm ext}}
\def\gi{{\bf g}_{\rm int}}
\def\tdyn{{\rm T}_{\rm dyn}}
\def\rp{r_{\rm P}}
\begin{document}
\title{The dynamical phase transitions of stellar systems and the corresponding kinetimacs}
\author[]{Xufen Wu$^{1}$, Pavel Kroupa$^{1,2}$\\
$^{1}$ Argelander-Institut fuer Astronomie der Universit\"{a}t Bonn, Auf dem H\"{u}gel 71, D-53121 Bonn, Germany \\
$^{2}$ Helmhotz-Institit f\"{u}r Strahlen-und Kernphysik, Universit\"{a}t Bonn, Nussallee 14-16, D-53115 Bonn
}

\maketitle

\begin{abstract}
External fields in Migromian dynamics \citep[MD or MOND,][]{Milgrom1983a} break the Strong Equivalence  Principle (SEP) and change the dynamics of self-bound stellar systems moving in space-varying background gravitational fields. We study two kinds of re-virialisation of the stellar systems: the violent phase transition and the adiabatic phase transition for systems moving on radial orbits, where the external field evolves from strong to weak and whose corresponding dynamics changes from Newtonian to Milgromian. We find that the time scale for the phase tranformation from Newtonian to Milgromian gravity lies only within one to a few crossing times for low density globular clusters with masses ranging from $10^4\msun$ to $10^6\msun$. Thus a globular cluster can appear frozen in the Newtonian regime despite being in the Milgromian regime for not longer than a few crossing times.
We also study the kinematics and anisotropy profiles of the systems. The velocity dispersions of the systems are larger after the phase transitions, especially for the outer regions of the stellar systems. Moreover, the isotropic systems become radially anisotropic, especially for the outer parts, after the process caused by the dynamical phase transition. Deeper Milgromian systems have more radially anisotropic velocity dispersion functions. We also find that the final profiles of density, velocity dispersion and anisotropy do not depend on the details of the phase transition. I.e., the mass distribution and kinematics of the end-states of the globular clusters do not depend on the rapidity of the transition from Newtonian to Milgromian gravity. Thus, the transition from the Newtonian to the Milgromian regime naturally induces a significant radially anisotropic velocity distribution in a globular cluster.
\end{abstract}
\section {Introduction}
Milgrom's dynamics \citep[hereafter MD,][]{Milgrom1983a} links the gravity in galaxies with the baryonic distribution, without Cold or Warm Dark Matter. The original Bekenstein-Milgrom's equation \citep{BM1984} is fully predictive on galactic scales: the baryonic Tully-Fisher relation \citep{TF1977}, the shapes of rotation curves of low and high surface brightness galaxies \citep{Milgrom_Sanders2003,Sanders_Noordermeer2007}, the dark matter effect in tidal dwarf galaxies \citep{Gentile_etal2007}, a universal scale of baryons and dark matter at the core radius of effective dark matter \citep{Gentile_etal2009}, the faster rotation speeds in polar rings \citep{Lughausen_etal2013}. Besides the success on the galactic scale, a cosmological model based on MD with massive sterile neutrino was proposed recently \citep[][]{Angus2009,Angus_Diaferio2011}, which matches the matter power spectrum, forms the right order of magnitude number of X-ray clusters and forms the high relative speeds of pairs of halos like the Bullet Cluster.
It is important to keep MD as a realistic option for small scale gravitational astrophysics because the Newtonian approach is problematic \citep{Kroupa_etal2010,Kroupa2012,Kroupa_etal2012}. MD is essentially a dark matter theory with $100\%$ conspiracy of phantom dark matter with baryons: the baryons dictate via the field equation how much phantom dark matter should be in a stellar system \citep{BM1984,Famaey_etal2007,Wu_etal2007},
\bey\label{poisson}
-\nabla &\cdot& \left[ \mu(X) ({\bf g}_{\rm ext} - {\mathbf\grad} \Phi_{\rm int}) \right]=4\pi G\rho,\\
\qquad X&=&{|{\bf g}_{\rm ext} - {\mathbf\grad} \Phi_{\rm int}| \over a_0} .\nonumber
\eey
Here $a_0$ is Milgrom's (1983) acceleration constant and we use $a_0=3.7~\pc \Myr^{-2}$. The above equation shows that the internal dynamical potential $\Phi_{\rm int}$ of a gravitating system in MD depends on the baryon density $\rho$ and its acceleration $\ge$ in the external background field, even if the latter is constant and uniform. The interpolating function $\mu$ satisfies
\bey\label{asymptotic}
&\mu& \rightarrow X \,\,\,\,(X\ll 1),\nonumber \\
&\mu &\rightarrow 1 \,\,\,\,(X \gg 1).
\eey
This ensures that the internal gravity acceleration $\gi$ is Newtonian when $X\gg 1$ and is in the deep MD regime when $X\ll 1$. The external field $\ge$ truncates the logarithmic potential of an isolated Milgromian system and enables stars to escape. In contrast to the Newtonian case, because the dynamics of a system depends on both internal and external fields, the Strong Equivalence Principle (SEP) is violated in MD. The function $\mu$ can be derived from quantum mechanical process in space time \citep[][Appandix A]{Milgrom1999,Kroupa_etal2010}. Milgrom's dynamics is thoroughly reviewed by \citet{Famaey_McGaugh2011}.

The external field effect (EFE) has been studied for a variety of situations: the absence of dark matter in star clusters in the inner Milky Way disc \citep{Milgrom1983a}, the motion of probes in the inner solar system \citep{Milgrom2009, Iorio2010}, the Roche lobes of binary systems \citep{Zhao_Tian2006}, the escape speeds and truncations of galactic rotation curves \citep{Wu_etal2007}, the distant star clusters of the Milky Way \citep{Haghi_etal2011} and line-of-sight velocity dispersion of statellites surrounding a host galaxy \citep{Angus_etal2008}.

MD predicts that a self-bound gravitational system moving in a space-varying field should have varying internal dynamics, especially for diffuse systems. Stellar systems like low-central-density globular clusters and ultra-faint dwarf satellite galaxies are the best candidates to test gravitational dynamics. Those systems moving from the galactic centre to the outer parts should change their internal gravity from Newtonian to Milgromian. Therefore such systems are not in equilibrium when moving on such orbits, and there is an additional evolution of such systems. In this paper we are interested on the time scale the systems re-virialise, and their morphology and internal kinematics when the background field changes.

We construct models for low-central-density stellar systems with different parameters that are allowed by observations, and study their evolution during the violation of the SEP via: (1) violent process of the SEP violation by shifting the gravity suddenly from Newtonian to deep MD in \S \ref{sim}; (2) moving the systems on realistic radial orbits in the Milky Way Beson\c{c}on Milgromian potential \citep{Wu_etal2008} in \S \ref{real} and \S \ref{adiabatic}. We conclude our results in \S \ref{conc}.

\section{Extended Globular Clusters}\label{egc}
Globular clusters (GCs) are typically compact objects with half-light radii of a few pc and with masses in the range $10^4-10^6~\Msun$. The internal gravitational fields in the compact centres of normal GCs are usually much larger than Milgrom's acceleration constant $a_0$, hence the normal GCs are dominated by Newtonian gravity, and they are not good candidates to test the dynamics of alternative gravities. However there is a certain fraction of GCs \citep[$\approx 9\%$ for the Galactic GCs in][]{Harris1996} that have extended half-light radii of more than $10\pc$. The extended GCs are found widely existing around extragalactic systems and their half-light radii can reach up to $30~\pc$ \citep[][in preparation]{Larsen_brodie2000,Harris_etal2002,Chandar_etal2004,Lee_etal2005,Peng_etal2006,Chies-Santos_etal2007,Georgiev_etal2009,Huxor_etal2011,Bruns_Kroupa2011,Bruens_Kroupa2012}. In contrast to normal GCs, the extended GCs have low density centres, where the acceleration in the central regions can be below $a_0$. Therefore it is very important to test the dynamics of the extended GCs in different gravities. A test of the outer Galactic clusters is suggested by \citet{Baumgardt_etal2005}, with the assumption that the velocity dispersion profiles in the clusters are isotropic.

\begin{table*}
\begin{center}\vskip 0.00cm
\caption{Parameters of isolated mass models for extended globular clusters. The columns from left to right provide the following information of the models: models' ID ($1_{\rm st}$ column), total mass ($2_{\rm nd}$ column), Plummer radius $r_P$ ($3_{\rm rd}$ column), number of particles in the simulations ($4_{\rm th}$ column), crossing time of models in Newtonian gravity ($5_{\rm th}$ column) and in Milgromian gravity ($6_{\rm th}$ column), a characteristic time scale for the Milgromian Plummer model $r_P(GMa_0)^{-1/4}$ ($7_{\rm th}$ column), the stability parameter $\xi$ of the models after re-virialised ($8_{\rm th}$ column) and the corresponding colours used in Fig. \ref{checkg} ($9_{\rm th}$ column).}
\begin{tabular}{llllllllc}
\hline
Models ID & M($\msun$) & $\rp$ ($\pc$) & N & $\tcn~(\Myr)$ & $\tcm~(\Myr)$ & $r_P(GMa_0)^{-1/4}~(\Myr)$ & $\xi$& Colour\\
\hline
1 & $10^6$ & 10  & 100000 & 3.53 & 2.73 & 0.89  & 1.11 &Black\\
2 & $10^5$ & 10  & 100000 & 11.69 & 5.70 & 1.58 & 1.65 &Magenta\\
3 & $10^4$ & 10  & 25000  & 36.58 & 10.42 & 2.81 & 2.26 &Cyan\\
4 & $10^5$ & 5   & 100000 & 4.07 & 2.72 & 0.79 & 1.25 &Yellow\\
5 & $10^5$ & 20  & 100000 & 32.53 & 11.70 & 3.16 & 2.37& Green\\
\hline
\end{tabular}
\label{ics}
\end{center}
\end{table*}

\begin{figure}{}
\begin{center}
\resizebox{9.cm}{!}{\includegraphics{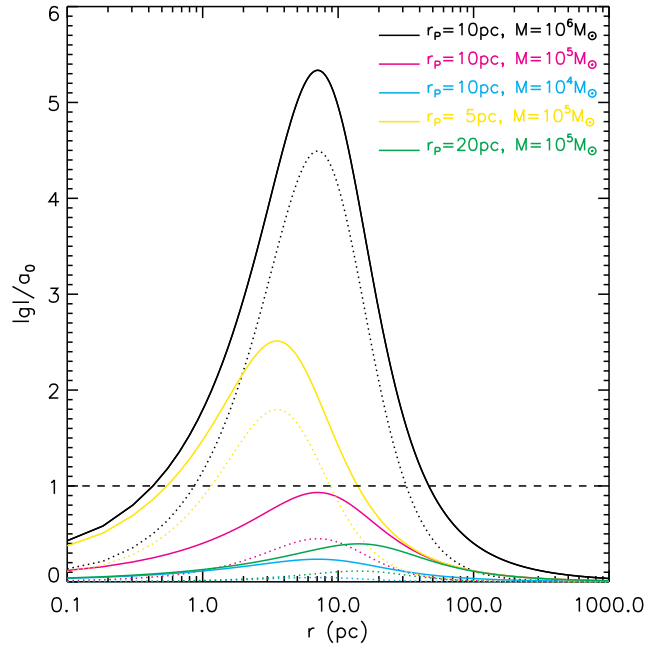}}
\makeatletter\def\@captype{figure}\makeatother
\caption{The gravitational acceleration for the models in Table \ref{ics}. The solid and dotted lines are acceleration from the Milgromian and Newtonian Poisson equations, respectively. The dashed line is $a_0=3.7\pc \Myr^{-2}$.}\label{checkg}
\end{center}
\end{figure}

We use Plummer's density profile \citep{Plummer1911,BT2008} to model the extended GCs,
\beq
\rho(r)=\frac{3M}{4\pi \rp^3}\left(1+\frac{r^2}{\rp^2}\right)^{-5/2},
\eeq
where $\rp$ is the scale radius and has a relation with half mass radius $r_h\simeq 1.3~\rp$, $M$ is the total mass. We use the method of \citet{Gerhard1991} to construct isotropic N-body Initial Conditions (ICs) in Newtonian gravity for the violent re-virialisation simulations. We summarise the parameters of the mass models in Table \ref{ics}, and the parameter $N$ in Table \ref{ics} is the number of N-body particles. The accelerations of the analytical mass models (not from the N-body particles) obtained from the Milgromian (Eq. \ref{poisson}) and Newtonian Poisson equations are shown in Fig. \ref{checkg}. We find that the mass models $2$, $3$ and $5$ are in the deep MD regime while model $1$ is mostly ($r < 5\rp$) in the Newtonian regime, and model $4$ is dominated by mild-Milgromian gravity.

\section{Violent phase transition}\label{sim}
The most direct way to study the phase transition in different gravities is virialising the Newtonian equilibrium models with the Milgromian Poisson equation (Eq. \ref{poisson}). A GC which is shot from the galactic centre to the outer regime, if the Galactic orbital time is much shorter than the re-virialisation time scale, experiences such a violent dynamical phase transition.
\subsection{Numerical setup}\label{num}
The interpolating function $\mu(X)$ has several popular forms giving the same asymptotic behavior in Eq. \ref{asymptotic}.
We shall apply the `simple'-$\mu$ function in the following of this paper \citep{Famaey_Binney2005,Sanders_Noordermeer2007,Wu_etal2007},
\beq
\mu(X)=X/(1+X).
\eeq
In order to solve the non-linear Poisson equation (Eq. \ref{poisson}), we use the particle-mesh N-body code NMODY \citep{nmody}. NMODY is developed for isolated Milgromian gravity systems and has been well tested \citep{Nipoti_etal2007,Nipoti_etal2008,Nipoti_etal2011}. It solves Newtonian potentials by the spherical harmonic expansion to the differential Poisson equation and then iterates into the Milgromian potential. For the simulations in this section, we choose a grid-resolution of $n_r\times n_{\theta} \times n_{\phi}=256 \times 32 \times 64$, where $n_r,~n_\theta,~n_\phi$ are the number of grid cells in radial, polar and azimuthal dimensions, respectively. The radial grids are defined as $r_i = r_s\times \tan \left[(i+0.5)0.5\pi /(n_r+1)\right]$ with $r_s=20\pc$ and $i=0,1,2,...,n_r$, and the angular grids are equally segmented. NMODY uses the leap-frog scheme to integrate the motions of the particles. The time steps are globally defined as $\dt= \frac{0.1}{\sqrt{\max |\nabla \cdot {\gi}|}}$, meaning that there are around $10$ time steps on an orbital loop for the particles in the densest regime. Therefore the time steps are small enough to avoid the artifical run-aways of the particles in the central regime.

In order to choose the correct time scale for our simulations, we define the crossing time of the models in this paper as $T_{\rm cross} = r_{\rm 90}/v_{\rm circ,r_{\rm 90}}$, where $r_{\rm 90}$ is the radius enclosing $90\%$ of the total mass of a model, and $v_{\rm circ,r_{\rm 90}}$ is the corresponding circular velocity.
In Newtonian dynamics, $v_{\rm circ,r_{\rm 90}}^{\rm Newt} \equiv \sqrt{GM_{90}/r_{90}}$, where $M_{90}$ is $90\%$ of the total mass, while $v_{\rm circ,r_{\rm 90}}^{\rm MD} \equiv (GM_{90}a_0)^{1/4}$ in Milgromian dynamics.
The crossing time is the time scale within which the majority of a system virialises. The values of $\tcn$ (i.e., crossing times of the Newtonian models) are listed in Table \ref{ics}. The crossing times of the models range from 3.5 to 36.7 $\Myr$. The corresponding two-body relaxation times are $t_{\rm rel}~\approx~\frac{0.1N}{\ln N} T_{\rm cross}~\approx~ 100~T_{\rm cross}$, such that two-body-encounter driven processes can be neglacted during the simulations.

Before shifting to Milgromian gravity, we first virialise the models contructed from \S \ref{egc} in Newtonian gravity for $\approx 25~\tcn$, to fully phase mix the systems and test the stability of the unperturbed systems. The details of the stability test can be found in Appendix \ref{stability}. Then we evolve the systems in Milgromian gravity for another $\approx 100-200~\Myr$ ($200~\Myr$ for models $3$ and $5$ to ensure there are more than $10~\tcm$ for each system) to study their re-virialisation process.
\subsection{Time scales}\label{vio_vir}
\subsubsection{Time scale of re-virialisation and the virial ratio}\label{ts_vir}
\begin{figure}{}
\begin{center}
\resizebox{9.cm}{!}{\includegraphics[angle=-90]{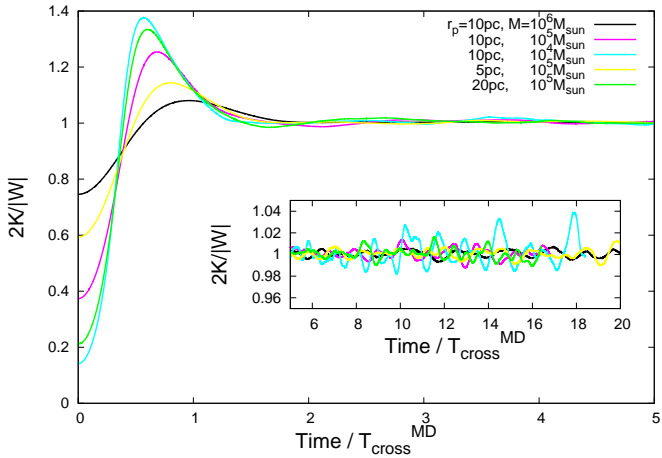}}
\makeatletter\def\@captype{figure}\makeatother
\caption{The virial ratio during the re-virialisation due to the phase transition. Time = 0 at the time when Newtonian gravity instantly transits into Milgromian gravity. After about 5 crossing times (see \S \ref{ts_vir} for details) in Milgromian gravity, the systems are in their new equilibrium states.}\label{vir_violent}
\end{center}
\end{figure}
For a collisionless system, the scalar virial equation should be satisfied if the system is in equilibrium \citep{BT2008}:
\beq
2K+W=0,
\eeq
 where $K$ is the kinetic energy of the system and $W$ is Clausius' integral, $W=\int \rho {\vec x} \cdot \nabla \Phi d^3x$ \citep[where ${\vec x}$ is the spatial vector,][]{Clausius1870}. \citet{Lynden-Bell1967} showed that for a Newtonian system violently virialising to equilibrium, the time scale is approximately ${3 T_r^*}/{8\pi} \simeq {T_r^*}/{8} $, where $T_r^*$ is the typical radial period of the system at the equilibrium radius. $T_r^*$ and $T_{\rm cross}$ should be of comparable magnitude. Therefore the violent virialisation time is comparable to a crossing time at the equilibrium radius. \citet{Nipoti_etal2007} studied dissipationless collapses of systems in MD and they obtained a virialisation time $\propto r(GMa_0)^{-1/4}$ for deep MD systems collapsing from rest (i.e. the initial velocities of the particle systems are zero). Since our systems evolve from one equilibrium state to another equilibrium state, the time scale for the re-virialisation should be comparable to this violent virialisation time scale in Milgromian gravity.

Fig. \ref{vir_violent} shows the time scale of the re-virialisation of the models. The virial ratios $\vir$ of all the models are smaller than $1$ when the gravity is switched to Milgromian at Time = 0. The particles are accelerated in the deepened potential. We use the definition of a system's `dynamical time' $\tdyn$ in \citet{Nipoti_etal2007}, i.e. the time scale when $\vir$ reaches its maximum value. $\tdyn$ is within $0.5-1.0~\tcm$.  
The varying trend of the time scale for different models agrees with the violent re-virialisation time scale in \citet{Nipoti_etal2007}. The $\tdyn$ of re-virialisation is approximately proportional to $ \rp(GMa_0)^{-1/4}$ here as well. We show the values of $ \rp(GMa_0)^{-1/4}$ in Table \ref{ics}. We define the re-virialisation time to be $5\times \tdyn$, since the amplitude of osscilation of $\vir$ around $1$ is smaller than $1.5\%$ at $T>5 \tcm$ in Fig. \ref{vir_violent}, and the systems can be considered fully re-virialised.
From Fig. \ref{vir_violent} we find that the time scale for the systems to re-virialise from Newtonian to Milgromian gravity is rather short, especially for the violent re-virialisation period.

\subsubsection{Lagrangian radii, local time scales and mass profiles}\label{rlag}
\begin{figure}{}
\begin{center}
\resizebox{8.7cm}{!}{\includegraphics[angle=-90]{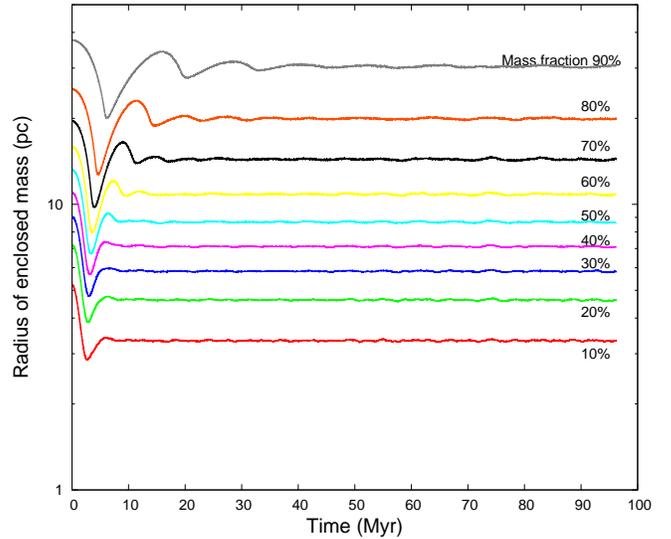}}
\makeatletter\def\@captype{figure}\makeatother
\caption{The evolution of the $10\% - 90\%$ Lagrangian radii for model $2$. At Time = 0, Newtonian gravity switches instantly to Milgromian gravity.}\label{rmass_violent}
\end{center}
\end{figure}

We study the evolution of mass profiles of the models by considering their Lagrangian radii. We show one example (model $2$) of the evolution of the $10\%$ to $90\%$ Lagrangian radii in Fig. \ref{rmass_violent}. We define the `dynamical time of each $10\%$ of mass shell' (short for fractional dynamical time $\tdyn^i$, $i=1,2,...9$.) as the time when its Lagrangian radius collapses to its minimal value. We find that the fractional dynamical time scales, $\tdyn^i$, are different for different enclosed percentage of mass: the time scale increases with increasing enclosed mass. The Lagrangian radii become smaller when re-virialised to Milgromian gravity especially for the inner regime. The $\tdyn^i$ are important since they show the maximum time a regime can remain frozen in Newtonian gravity.

\begin{figure}{}
\begin{center}
\resizebox{9.cm}{!}{\includegraphics[angle=0]{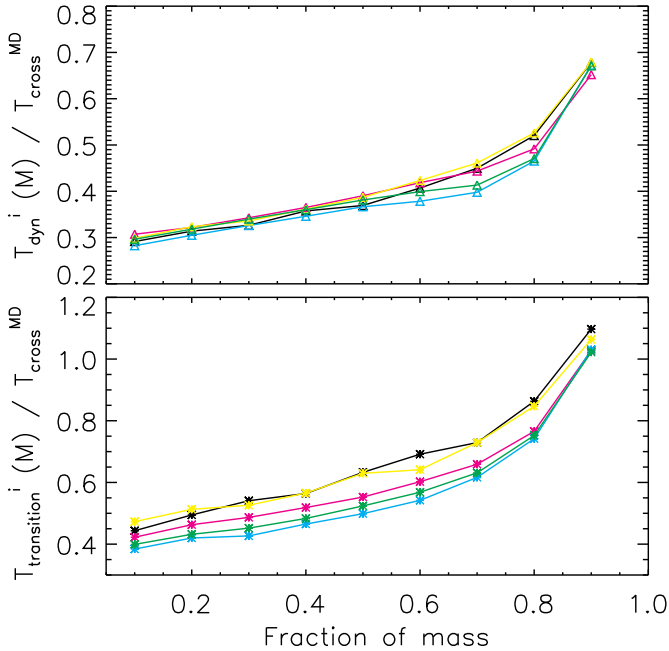}}
\makeatletter\def\@captype{figure}\makeatother
\caption{{\bf Upper panel:} The time scale of the phase transition in units of the MD crossing time at the $10\%-90\%$ Lagrangian radii. {\bf Lower panel:} $\tdyn$ at the $10\%-90\%$ Lagrangian radii. The colours are for different models, as defined in Fig. \ref{checkg}. }\label{tpt}
\end{center}
\end{figure}

We show the fractional dynamical times $\tdyn^i$ of all the models in the upper panel of Fig. \ref{tpt}. We find that $\tdyn^i$ (in unit of $1\tcm$) as a function of the mass fraction is almost the same for different models: from $0.3~\tcm$ to $0.7~\tcm$ for $10\%-90\%$ of the enclosed mass. The innermost parts of the systems are frozen in Newtonian gravity for times shorter than $0.3~\tcm$. The outermost parts of the systems are frozen for times shorter than $0.7~\tcm$.

For each mass shell, the first oscillation from the initial value to the first maximum of the Lagrangian radius should be approximately a local crossing time. We call the time for the first oscillation of Lagrangian radii in each shell the `local transition time', ${\rm T}_{\rm transition}^i$. ${\rm T}_{\rm transition}^i$ versus mass fraction has the same trend as $\tdyn^i$ in the upper panel. The local transition time scales are from $0.4~\tcm$ to $1.1~\tcm$ for $10\%-90\%$ of the enclosed mass. However, there is a deviation of the local transition time curves between the different models normalised by their own $1\tcm$: in the lower panel of Fig. \ref{tpt}, comparing models $1$ (black curve), $2$ (Magenta curve) and $3$ (Cyan curve), the more massive models have longer time scales; and comparing models $2$ (Magenta), $4$ (Yellow) and $5$ (Green), the models with smaller initial radii have longer time scales. I.e., the models with denser density distributions, where the gravity is stronger, have longer time scales in unit of $\tcm$; while the more diffuse models where their gravities are weaker, have shorter time scales in unit of $\tcm$. This is because after $T_{\rm dyn}^i$, the local particle structures over-collapse in Milgromian potential, and then oscillate back to the equilibrium state, and the diffuse models are closer to the deep MOND equilibrium after $T_{\rm dyn}^i$.

The evolution of Lagrangian radii implies that the mass profiles of the GCs have changed after re-virialisation. We show the spherically averaged density profiles of the models in the upper panel of Fig. \ref{prof_violent}. The densities, $\rho(r)$, are normalised by $\rho_0=M/\rp^3$ so that all the Newtonian models stay on the same curve. In general the central densities become larger and the core radii are smaller after the re-virialisation. This agrees with the evolution of the Lagrangian radii. We further note that the diffuse models change their central density more than the compact models. This is because the more diffuse models are further away from equilibrium when the gravity switches from Newtonian to Milgromian.

\subsubsection{Can Palomar 14-like GCs be frozen in Newtonian gravity?}
There are distant GCs like Palomar 14 (hereafter Pal 14), which is located in a deep MD background \citep[$74.7~\kpc$ distance to the Sun,][]{Hilker2006}, but appear to behave Newtonian with a small value of the velocity dispersion \citep{Haghi_etal2009}. This kind of GC seems to be a challenge to MD. However, \citet{Gentile_etal2010} argued that Pal 14 is not sufficient enough to falsify MD, and one of their arguments is that Pal 14 could be on a eccentric orbit around the Milky Way, and that its potential is Newtonian at its pericentre due to the strong external field from the Milky Way: the Newtonian-like potential is frozen when Pal 14 moves to the current position. From Figures \ref{vir_violent}, \ref{rmass_violent} and \ref{tpt} we find that the time scale of the transition is rather short and the systems go into the MD dominated regime in a few to a few tens of $\Myr$.
If the orbital time from the pericentre to the outer stellar halo of the Milky Way (i.e., far enough away where the external field is $\ll a_0$) is shorter than the `dynamical' time, it is possible that a GC may not be in equilibrium and its potential appears frozen in the Newtonian regime. We will briefly show the kinetics of rapid phase transition later in this section and show simulations for GCs moving on different radial orbits around the Milky Way in \S \ref{real}, for comparing the realistic time scale of re-virialisation and the orbital time.

\subsection{Kinematics}\label{text_kin}
\begin{figure}{}
\begin{center}
\resizebox{8.7cm}{!}{\includegraphics[angle=-90]{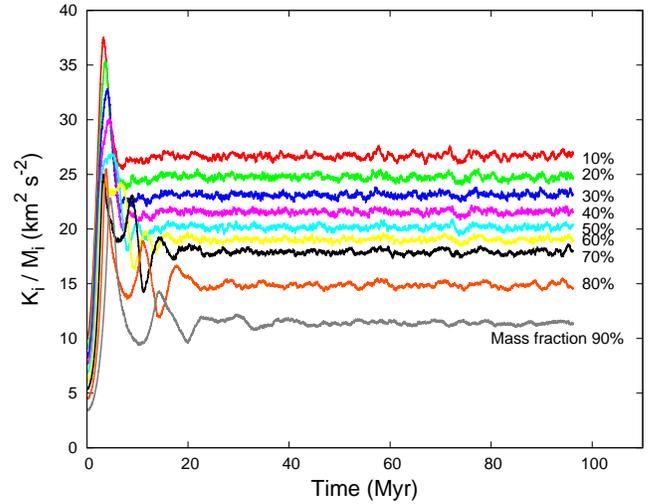}}
\makeatletter\def\@captype{figure}\makeatother
\caption{The evolution of the kinetic energy within shells of the $10\% - 90\%$ Lagrangian radii of model $2$. $M_i$ is the mass within the $i_{th}$ equal mass shell. The Newtonian gravity switches instantly into Milgromian gravity at Time = 0.}\label{rkin_violent}
\end{center}
\end{figure}

For the particles in each shell of the $10\%-90\%$ Lagrangian radii, the kinetic energy is defined as,
\beq
K_i=\frac{1}{2}\sum_{\rm ip=1}^{\rm N_{\rm parts,i}} m_{\rm ip} v_{\rm ip}^2,~~~~(i=1,2,...,9),\nonumber \eeq
where ${\rm N}_{\rm parts,i}$ is the number of equal mass particles in the $i_{th}$ shell and $m_{\rm ip}$ is the mass of ${\it ip}_{th}$ particle. We show the evolution of $K_i$ of model $2$ in Fig. \ref{rkin_violent}. The time scale of the first oscillation at different Lagrangian radii agrees with Fig. \ref{rmass_violent}. Each shell is fully virialised after about $5~\tdyn^i$. The kinetic energy increases by a factor of $2.7$ in the innermost shell and by a factor of $3.3$ in the outermost shell for this model.

\begin{figure}{}
\begin{center}
\resizebox{9.cm}{!}{\includegraphics{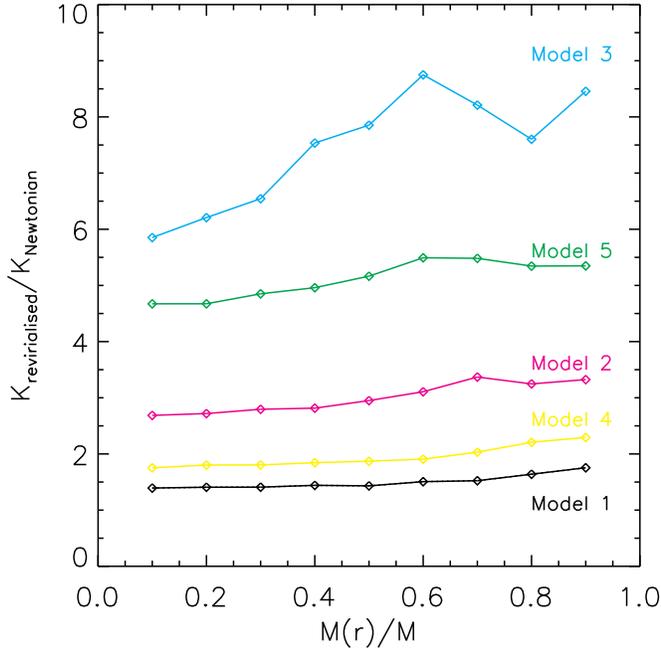}}
\makeatletter\def\@captype{figure}\makeatother
\caption{The increase of kinetic energy in each shell from Newtonian to Milgromian gravity. I.e., $\frac{K_{i,~{\rm re-virialised}}}{K_{i,~{\rm Newtonian}}}$. The colours are for different models, as defined in Fig. \ref{checkg}.}
\label{rkin_increase}
\end{center}
\end{figure}

We also study the increment of $K_i$ of all the models in Fig. \ref{rkin_increase}, i.e., $\frac{K_{i,~{\rm re-virialised}}}{K_{i,~{\rm Newtonian}}}$ versus the fraction of mass. We find that the increasing curves of kinetic energy are mildly increasing with the fraction of mass for all the models, i.e., the kinetic energy in the inner parts of the systems increase slightly less than the increment of their outer parts. The trends of the curves for all the models are rather similar. There is a large oscillation for the cyan curve (i.e., model $3$) since there are fewer particles in this model, and the particle noise is larger in this simulation.

\begin{figure}{}
\begin{center}
\resizebox{9.cm}{!}{\includegraphics{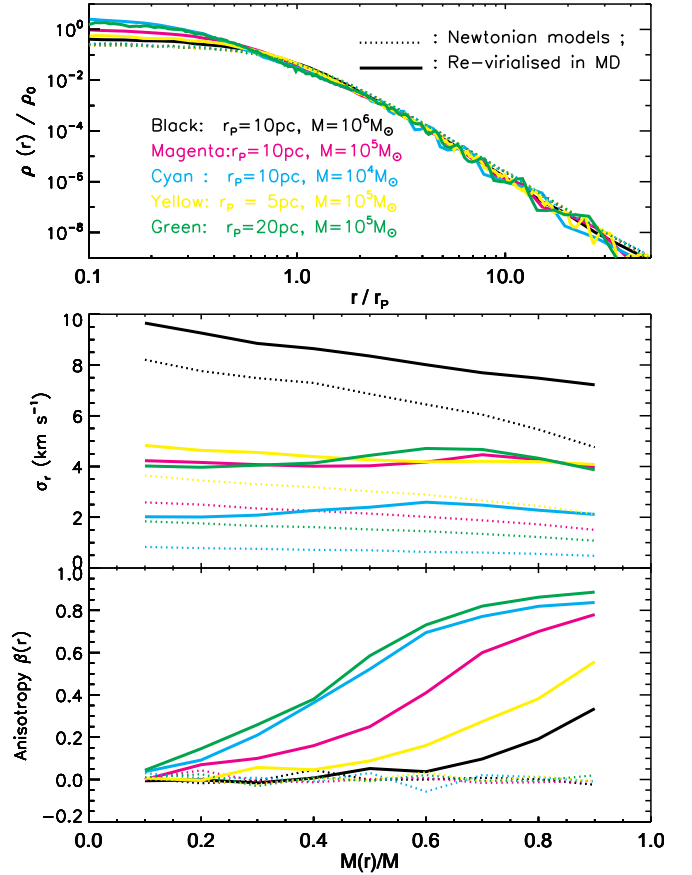}}
\makeatletter\def\@captype{figure}\makeatother
\caption{The spherically averaged density profiles ({\bf upper panel}), radial velocity dispersion ({\bf middle panel}) and anisotropy ({\bf lower panel}, Eq. \ref{beta}) for the models in Table \ref{ics}. The  dotted and solid lines correspond to, respectively, models in Newtonian and in Milgromian dynamics.}\label{prof_violent}
\end{center}
\end{figure}

The kinetic energies of the systems evolve significantly after the
re-virialisation. Therefore the velocity dispersions should also
evolve significantly. We study the radial velocity dispersion and
the corresponding anisotropy profiles of the models before and after
the re-virialisation. The middle panel of Fig. \ref{prof_violent}
shows the evolution of the $\sigma_r(r_i)$ profiles:
\beq\label{sigmar} \sigma_r^2(r_i) = \frac{1}{N_{\rm
parts,i}}\sum_{\rm ip=1}^{\rm N_{\rm parts,i}} (v_{r,{\rm ip}}-{\bar
v_{r,i}})^2,\eeq where $r_i$ is the $(i\times 10)\%$ Lagrangian radius, $v_{r,{\rm ip}}$ is the radial velocity of the ${\it ip}_{th}$ particle
and ${\bar v_{r,i}}$ is the mean radial velocity of the particles in the
$i_{th}$ equal mass shell. The original Newtonian models have
decreasing $\sigma_r(r_i)$ profiles (dotted curves) versus
increasing radii $r_i$, while the final products can have decreasing
profiles for the compact systems like models $1$ (black) and $4$
(yellow), or have $\sigma_r(r_i)$ profiles that increase mildly at
first and then decrease again in the outer regime for the loose
systems like model $2$. With a same mass (models $2$, $4$ and $5$,
i.e., magenta, yellow and green curves), the peak values of the
increasing $\sigma_r(r_i)$ profiles appear at a smaller mass
fraction for the models with a larger initial Plummer radius $r_{\rm
P}$; however the amplitudes of the $\sigma_r(r_i)$ profiles do not
change significantly with radius. This is different to the Newtonian
case. The $\sigma_r(r_i)$ profiles of self-consistent Newtonian
models (the dotted curves) are very different if their $r_{\rm P}$
are different.

Since the systems collapse during the re-virialisation, the orbital
structures might completely change. \citet{Nipoti_etal2007} showed
the anisotropy profiles of their rest models collapsing in Newtonian
and Milgromian gravities, and their models become highly radially
anisotropic after the virialisation, especially for their outer
parts where $r>3r_h$ ($r_h$ being the 3-dimensional half mass
radius) for their deep Milgromian model, and $r>r_h$ for
mild-Milgromian and Newtonian models. Although our ICs are different
to the ones in \citet{Nipoti_etal2007}, the re-virialisation from
the Newtonian to the Milgromian regime is similar to the collapse
process. We expect to obtain radially anisotropic models from our
rapid phase transition as well. The anisotropy profiles
$\beta(r_i)$ are defined in the same way as that in \citet{BT2008},
\beq\label{beta} 
\beta(r_i)\equiv
1-\frac{\sigma_\theta^2(r_i)+\sigma_\phi^2(r_i)}{2\sigma_r^2(r_i)},
\eeq 
where $\sigma_\theta$ and $\sigma_\phi$ are polar and
azimuthal components of velocity dispersion, and the definition is
similar to that of radial velocity dispersion in Eq. \ref{sigmar}.
We show the $\beta(r_i)$ profiles of our final products in the
lower panel of Fig. \ref{prof_violent}. Indeed all of the
$\beta(r_i)$ profiles are radially anisotropic. The more compact
models have larger anisotropic radii (i.e., the radii where the
models start to be anisotropic) and they are less radially
anisotropic, while the more diffuse systems have smaller anisotropic
radii and they are more radially anisotropic. The $\beta(r_i)$
profiles of model $3$ and $5$ are remarkably radial, they are almost
linearly increasing from the $10\%$ to the $60\%$ enclosed mass
radii and have $\beta > 0.8$ at radii containing $60\%$ of the
enclosed mass, and then the $\beta(r_i)$ profiles mildly increase
with mass. Therefore diffuse models, whose self-gravity are in deep
MD, change the anisotropy profiles more significantly than the
compact models.

\subsection{Phase space distribution during re-virialisation}

\begin{figure}{}
\begin{center}
\resizebox{9.cm}{!}{\includegraphics{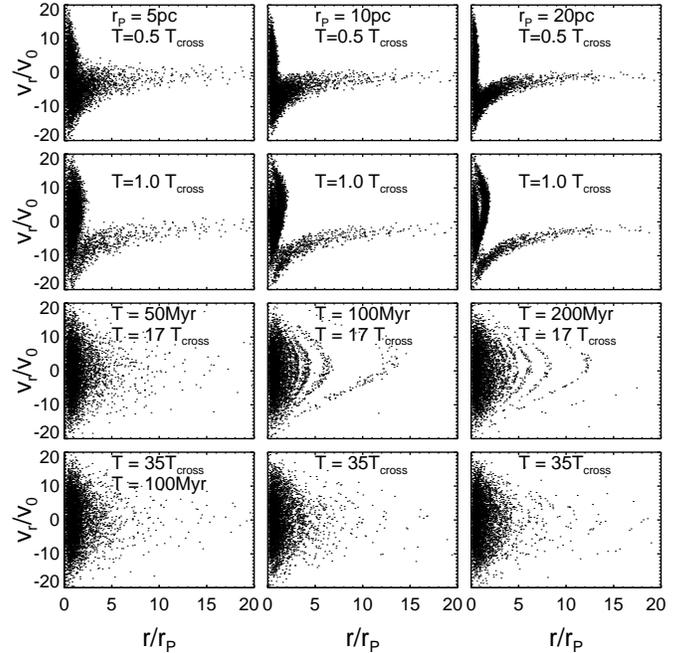}}
\makeatletter\def\@captype{figure}\makeatother
\caption{The phase space distribution at different snapshots of violent re-virialisation for models $4$ (left panels), $2$ (middle panels) and $5$ (right panels). The radii are scaled by the initial Plummer radius $r_{\rm P}$, and the radial velocities of particles, $v_r$, are scaled by $v_0=\sqrt{GM/r_{\rm P}}$.}\label{ps_violent}
\end{center}
\end{figure}

In order to quantitatively compare the re-virialisation of
systems from Newtonian to Milgromian dynamics with
\citet{Nipoti_etal2007}'s dissipationless collapse process, the
phase space distributions of the models are studied and three
typical examples are shown in Fig. \ref{ps_violent}: models $4$
(left panels), $2$ (middle panels) and $5$ (right panels), which
represent systems in weak-, moderate- and deep-Milgromian dynamics,
respectively. The phase space distributions are studied at times of
$0.5~\tcm$, $1~\tcm$, $17~\tcm$ and $35~\tcm$. The dynamical times
for the majority parts of the systems are shown in Fig. \ref{tpt},
and $1~\tdyn$ roughly amounts to $0.7~\tcm$ for the majority parts
($90\%$ enclosed mass) of the systems. Therefore, $35~\tcm$ roughly
amounts to $50~\tdyn$. The times at which the phase space
distributions are studied here are analogous to those in \citet[][
$0.5~\tdyn$, $1.0~\tdyn$, $44~\tdyn$]{Nipoti_etal2007}. Within
$35~\tcm$ the systems are fully re-virialised, although in the very
outer regimes there are few particles which have not yet phase mixed. We have to note that the amount of such particles are
actually small and that there are only about $1.5\%$ of mass outside
a radius of $10~r_{\rm P}$.

At the beginning of violent re-virialisation ($0.5~\tcm$ and
$1~\tcm$), particles fall into the centres of the systems, and some of the particles have
already crossed the centres, which correspond to the particles near
$r~=~0$. This is similar to that of dissipationless collapse.
However, the phase space distributions have wider spreads compared
to that at the beginning of dissipationless collapse (Nipoti et al.
2007). The reason is that the phase transitions of our models start
from Newtonian equilibrium state, while the dissipationless collapse
processes start from rest. At time $T~=~1.0~\tcm$, the deeper the
Milgromian dynamics dominating the systems is, the clearer the
shell-like structures appear near the centre of the systems. Those
shell-like structures are particles moving in and out of the systems
in the centre regimes. Before the systems are phase mixed, the
shell-like structures will appear in the outer regimes of the
systems as time is proceeding. At $T~=~17~\tcm$, the weak-Milgromian
system has already erased all structures. This implies that it approaches a phase mixed state.
However, the other two models still have shell-like structures
outside $3~r_{\rm P}$, especially for the deep-Milgromian system.
This is similar to the case of dissipationless collapse since the
Milgromian system is less efficient in phase mixing. The enclosed
mass within $3~r_{\rm P}$ is $85.4\%$ of the overall mass.
Therefore, the regimes which contain shell-like structures at
$T~=~17~\tcm$ are the outer regimes of the systems. We further
evolve the systems $2$ and $5$ until $T~=~35~\tcm$, and then compare
with model $4$. After $35~\tcm$ the shell-like structures of all the
models disappear, which means the models are fully phase mixed. The
phase mixing process in the re-virialisation case is shorter than that in the
dissipationless collapse process \citep{Nipoti_etal2007}.

\subsection{Radial Instability after systems re-virialised}

From the studies of \S \ref{text_kin}, one can see that the
diffuse systems going into the deep-Milgromian dynamics regime (Models $3$
and $5$) are strongly radially anisotropic from the half-mass radii outwards.
There are many radial orbits in the deep-Milgromian systems after
re-virialisation. Since the re-virialisation of the systems is a new
prediction compared to Newtonian dynamics, it is important to
examine whether the re-virialised systems are stable, i.e., whether
their velocity dispersions become isotropic by the effect of
radial instability.  A parameter describing the ratio between radial
and tangential anisotropies was defined in
\citet{Trenti_Bertin2006,Polyachenko_Shukhman1981},
\beq\label{xi}
\xi=\frac{2K_r}{K_t},\eeq
for a globle
(Newtonian) system, to characterize the stability of the system. Here $K_r$ and $K_t=K_\theta+K_\phi$ are the radial and tangential components of the kinetic energy tensor, respectively. $K_r\equiv \sum_{ip=1}^{N}\frac{1}{2}m_{\rm ip}v_{r,{\rm ip}}^2$, where $N$ is the total number of particles for a system. The polar and azimuthal components of the kinematic energy tensor, represented by $K_\theta$ and $K_\phi$ respectively, can be defined in the same manner.
\citet{Nipoti_etal2011} applied the $\xi$ parameter for Milgromian
systems, and found that there is an empirical stability criterion
for Milgromian systems: $2.3\le \xi_c \le 2.6$ for the stellar
components, where the $\xi_c$ is the critical value of $\xi$ which
depends on the stellar density and internal gravitational
acceleration. The systems are stable if $\xi \le \xi_c$. The $\xi$
values for the models after re-virialisation are listed in the
$8_{\rm th}$ column of Table \ref{ics}. The values of $\xi$ for
models $1$, $2$, $3$ and $4$ are smaller than $2.3$, and $\xi$ for
model $5$ is 2.37, which is within the range of $\xi_c$. Therefore
models $1-4$ are stable models, and model $5$ is at the critical
limit.

\section{Globulars in the space-varying gravitational field of the Galaxy}\label{real}
We have already studied the violent re-virialisation in \S \ref{sim}, and we showed that the re-virialisation takes only a few crossing times in Milgromian gravity, which is too short to freeze the distant GCs in Newtonian gravity. However a real star cluster moving in the Galactic potential experiences gradually evolving dynamics. The self-gravity of an equilibrium system moving in such a field is gradually evolving, and the transition between (quasi-) Newtonian and Milgromian gravity may be adiabatic. Thus an adiabatic contraction or expansion process is expected for such a system.
However this has not yet been studied for a system moving in a gravitational background field which is continuously space-varying.

Here we are interested in the adiabatic contraction process since it is a brand new physical process. %
We will select and virialise the Newtonian systems from \S \ref{egc} in a strong external field from the Milky Way in \S \ref{prekick}, and then send the systems on different radial orbits along the vertical $z$-axis of the Milky Way in \S \ref{orbits}. We shall compare the time scale of the phase transition, the kinematics and mass profiles of the final systems in \S \ref{adiabatic}.
\subsection{Background gravitational fields from the MW}\label{select}
The internal dynamics of a system embedded in an external field has been studied by \citet{Zhao_Tian2006} and \citet{Wu_etal2007}. The Possion equation of this sytem is linarised in the radii where the system is dominated by the external field. Assuming the external field is along the $z$-axis and using $(x,~y,~z)$ Cartesian coordinates,
\beq\label{constant}
\Phi_{\rm int}^{\infty} (x,y,z) = -\frac{GM}{\mu_e \sqrt{(1+\Delta_e)(x^2+y^2)+z^2}},
\eeq
where $\Phi_{\rm int}^{\infty}$ is the internal potential at infinity (i.e., at radii of the internal system where the self-gravity $\gi$ is much smaller than $\ge$, we can ignore $\gi$ in Eq. \ref{poisson}), $\mu_e$ is the $\mu$ function of the external field $\ge$, and $\Delta_e \equiv \frac{d \ln \mu_e}{d \ln X_e}|_{X_e=|\ge|/a_0}$.

Another important issue is the tidal field of the background. In \citet{Zhao_Tian2006} the tidal radii of binary systems in MD-like gravity have been studied,
\beq\label{rt}
r_{\rm tidal}=\left[\frac{M_{\rm int}}{(1+\zeta)M_{\rm ext}}\right]^{\frac{1}{3}}D_0,
\eeq
where $D_0$ is the orbital distance of the internal system, $M_{\rm int}$ is the total mass of the internal system and $M_{\rm ext}$ is the mass of the external system enclosed within the radius $D_0$, $\zeta=-\frac{d\ln g_{\rm int}}{d\ln D_0}=1-\frac{d\ln v_{\rm cir}^2}{d\ln D_0}$, where $\zeta=1$ in deep MD gravity and $\zeta=2$ in Keplerian (or Newtonian) gravity, and $v_{\rm cir}$ is the circular velocity at the radius $D_0$. For radii $r<r_{\rm tidal}$ of the internal system, the tidal field is not important, and the background field is dominated by a homogeneous external field. A numerically homogeneous external field can be added as a boundary condition while solving the Poisson equation \citep{Wu_etal2007,Wu_etal2008}. The same boundary conditions were introduced into NMODY in \citet{Wu_etal2010}.

Since Plummer profiles are extended density profiles, it is impossible and unnecessary to study $100\%$ of the enclosed mass. We define the inner $90\%$ of the total mass as the majority of the system. We only focus on the dynamics of the inner $90\%$ of the system, i.e., for $r\simeq 3.7~\rp$. Therefore only if $r_{\rm tidal} > 3.7~\rp$, the homogeneous boundary condition can be applied.

We shall model GCs moving in the Milky Way's potential. The Milgromian Besan\c{c}on Milky Way model \citep{Robin_etal2003} is studied by \citet{Wu_etal2007,Wu_etal2008} and \citet{Bienayme_etal2009}, in which the dark matter halo is removed and Milgromian dynamics is applied. We reproduce the Milky Way's Milgromian potential and gravity with the Besan\c{c}on density profile \citep{Robin_etal2003,Wu_etal2007,Wu_etal2008} in this work, using the code called NMODY \citep{nmody}, with a resolution of $n_r^{\rm MW}\times n_{\theta}^{\rm MW} \times_{\phi}^{\rm MW}=500 \times 64 \times 128$ using spherical coordinates $(r,~\theta,~\phi)$. The method of grid segmentation is the same as in \S \ref{num}, and $r_s^{\rm MW} = 10.0\kpc$. The angular resolution of the spherical harmonic expansion is $l_{\max}^{\rm MW}=16$. There is a weak external field of $0.01~a_0$ applied to the MW in the direction Sun-Galactic centre, which comes from the combination of the local gravitational attractors, i.e., the Great Attractor \citep[see][]{Radburn-Smith_etal2006}, the M31 galaxy, the Coma and Virgo clusters \citep[more details of the modelling of the Galaxy are in][]{Wu_etal2007,Wu_etal2008}.

\begin{figure}{}
\begin{center}
\resizebox{9.cm}{!}{\includegraphics[angle=0]{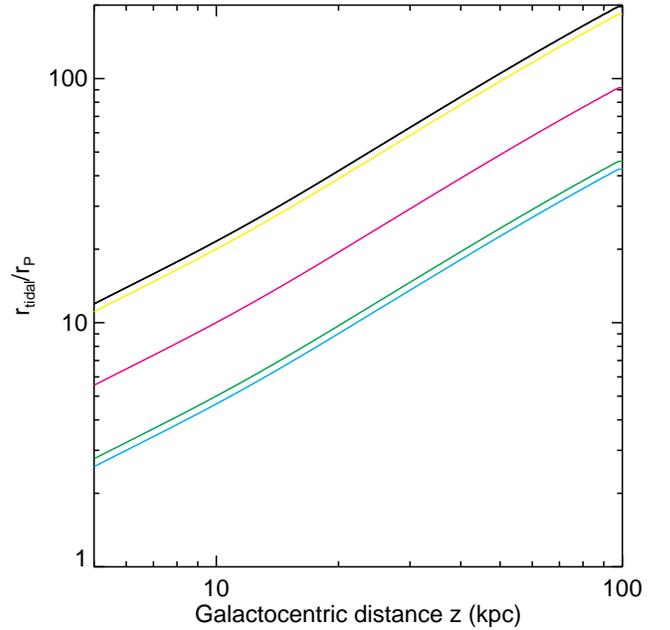}}
\makeatletter\def\@captype{figure}\makeatother
\caption{The tidal radii of the models from Table \ref{ics} at different Galactocentric distances. The Milgromian Besan\c{c}on Milky Way potential is used here. The colours for different models are the same as defined in Fig. \ref{checkg}.}\label{rtidal}
\end{center}
\end{figure}
We will move the systems along the vertical $z$-axis of the MW. We calculate the tidal radii of the models from Tab. \ref{ics} at different Galatocentric distances, from $5\kpc$ to $100\kpc$. The background field of the MW  at the position $(x,~y,~z)=(0,~0,~5)\kpc$ is $\approx 2.1a_0$, which is strong enough to reduce the Milgromian effect, and the dynamics of a GC embedded in this external field is Newtonian-like. We show the tidal radii in Fig. \ref{rtidal}. Note that the tidal radii of models $3$ and $5$ (cyan and green curves) are only $\approx 2.5 \rp$ when they are placed at the position $(x,~y,~z)=(0,~0,~5)\kpc$. Therefore models $3$ and $5$ are not suitable to use the homogeneous boundary condition in Eq. \ref{constant}. Theoretically we can choose a larger Galatocentric distance, say $z=10\kpc$, as starting points of the moving orbits for models $3$ and $5$. However the background field from the MW at $(x,~y,~z)=(0,~0,~10)\kpc$ is also weaker, $\approx 1.0a_0$. The Milgromian effects are remarkable for models $3$ and $5$ in such an external field. 
Therefore models $1$, $2$ and $4$ in Table \ref{ics} are more interesting. \footnote{For comparison, we will also show the re-virialisation of models $3$ and $5$ at $(x,~y,~z)=(0,~0,~10)\kpc$ and then will move them on radial Galactic orbits.}
\begin{figure}{}
\begin{center}
\resizebox{9.cm}{!}{\includegraphics{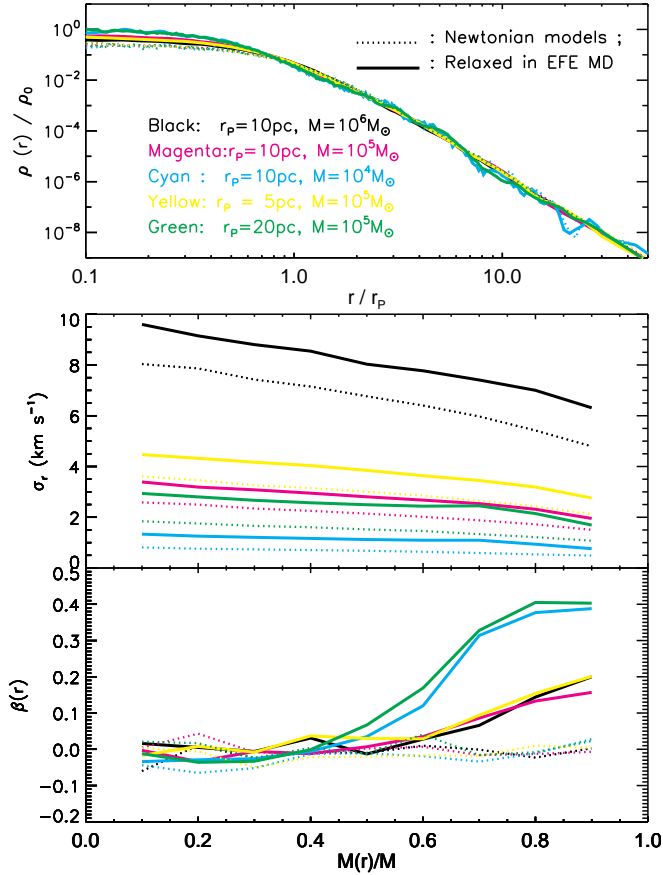}}
\makeatletter\def\@captype{figure}\makeatother
\caption{The spherically averaged density (upper panel), radial velocity dispersion (middle panel) and anisotropic profiles (lower panel, Eq. \ref{beta}) for models $1$ (black), $2$ (magenta) and $4$ (yellow) before (dotted curves) and after (solid) virialisation in strong Galactic fields. The models $3$ (cyan) and $5$ (green) are overplotted as comparison.}\label{relax}
\end{center}
\end{figure}

\subsection{ICs in quasi-Newtonian dynamics}\label{prekick}
At the Galatocentric position of $(x,~y,~z)=(0,~0,~5)\kpc$, the external field from the MW is strong. Therefore GCs staying at this position are dominated by quasi-Newtonian dynamics, i.e., $\mu(X)=\mu(\frac{|\ge|+|\gi|}{a_0})\approx 1$ in Eq. \ref{poisson}. The boundary conditions of the Poisson equation follow Eq. \ref{constant}. The differences to Newtonian dynamics are: there is a constant factor of $1/\mu_e$ on depth of the potential and a dilation factor of $(1+\Delta_e)$ on the shape of the potential. Therefore the quasi-Newtonian potential is $(1/\mu_e-1)$ deeper and it is prolate. Thus it is necessary to virialise the systems in the quasi-Newtonian potentials first, to ensure the equilibrium of the systems.

We virialise the models of interest in the background field of
${\ge}=(0,0,2.1a_0)$ for about $100~\Myr$. We show the density,
radial velocity dispersion and anisotropy profiles in Fig.
\ref{relax}. We find that after the virialisation, the density
profiles do not change significantly, however the inner regime of
the GCs become denser, by a factor of $1-3$. The radial velocity
dispersion profiles are systematically shifted, with a $20\%$
increment. The shape of the velocity dispersion profiles do not
change after the virialisation. The change of density and radial
velocity dispersion are due to the deepening of the potential. We
also find that the isotropic models become slightly radially
anisotropic after the virialisation (the lower panel). The systems
become radially anisotropic from the half mass radii of the systems
outwards and the anisotropies reach up to $0.2$ at the radii where
$90\%$ of the mass is enclosed. The radial anisotropies come from
the asymmetric internal potential due to the homogeneous external
field effect \citep{Wu_etal2010}: at the radii where the internal
and external fields are comparable, the potentials of the internal
systems are lopsided. The lopsidedness of the potential changes the
orbits in the systems. This is quite similar to the radial
anisotropy induced by tidal fields at an early stage in Newtonian
dynamics. There are already many contributions on the dynamical evolution of
star clusters in tidal fields in Newtonian gravity
\citep{Giersz_Heggie1997,Takahashi_Lee2000,Baumgardt_Makino2003,Lee_etal2006}.
In Newtonian dynamics, the tidal fields will bring in a strongly
radial anisotropy in the outer regimes of the star clusters at an early
stage, whereas the star clusters are still isotropic in the inner
regimes \citep{Takahashi_Lee2000}. The radial anisotropy in the
outer parts disappears quickly with time in realistic tidal
fields \citep{Baumgardt_Makino2003}. Finally, the outer regimes of
star clusters in Newtonian tidal fields become tangentially
anisotropic \citep{Giersz_Heggie1997,Lee_etal2006}. 

We have to note that this is not a tidal effect in MD, but an EFE which plays a similar role. The tidal radii (Eq. \ref{rt}) are much larger than the radii at which EFE becomes important. Comparing Fig. \ref{relax} with Fig. \ref{prof_violent}, we find that the anisotropy introduced by the EFE is milder than by the phase transition. However for the most compact system, model $1$, the anisotropy introduced by the two effects are comparable. This is because model $1$ is so compact that it is mostly dominated by Newtonian dynamics, and it does not evolve as much as the other more diffuse models during the phase transition.

In Fig. \ref{relax} we also virialise models $3$ and $5$ in the background field at the Galatic position of $(x,~y,~z)=(0,~0,~10)\kpc$ for comparison. The external field is as strong as $1.0~a_0$. We overplotted the revirialised initial Newtonian models in this Milgromian external field (cyan and green curves in Fig. \ref{relax}). We find that the density profiles and anisotropy profiles of models $3$ and $5$ change more compared to the other models: The centre densities are more concentrated and the anisotropy in the outer regimes is more radial for models $3$ and $5$, since at $(x,~y,~z)=(0,~0,~10)\kpc$ the Galactic field is weaker and the deviation from Newtonian dynamics becomes important. With a Galactic gravitational field of $1.0~a_0$, the Newtonian and Milgromian accelerations are comparable. 

\subsection{Radial orbits of GCs}\label{orbits}
The models are fully virialised for about $100~\Myr$ in the strong field in Milgromian gravity (when $a\gg a_0$ this becomes identical to Newtonian gravity). We shall move the virialised models ($1,~2$ and $4$) in the Galactic potential from near the Galactic centre, i.e., from a position at $(x,~y,~z)=(0,~0,~5)\kpc$. The external field changes fastest along the polar direction on a pure radial orbit for a given initial velocity. Besides, the major part of a system should be enclosed within its tidal radius, so we need to avoid GCs moving near the Galactic disc plane. Therefore we choose orbits along the polar direction and there are only non-zero values of the initial velocity along the $z$-axis.

We setup a grid resolution of $n_r^{\rm GC}\times n_{\theta}^{\rm GC} \times_{\phi}^{\rm GC}=100 \times 32 \times 64$, where the radial scaling parameter $r_s^{\rm GC}=20\pc$ and $l_{\max}^{\rm GC}=6$ since the GC models are spherically symmetric. The time steps are defined as in \S \ref{num}. In each simulation, we solve the Poisson equation for the Milky Way and store the gravitational acceleration field and potential on the grid, and then we interpolate the Galactic acceleration field and potential to the point where the centre of mass (CoM) of the GC is. Applying the boundary conditions of Eq. \ref{constant} onto the GC, the Poisson equation for the GC is computed. The positions and velocities of particles of the GC are updated each time step by the Leap-frog scheme, as in \S \ref{num}. The new CoM of the GC is calculated at the end of the time step, and then the external field of the new CoM of the GC is interpolated from the MW acceleration field and potential.
At each time step of the GC, the EF is updated from the CoM position, such that the GC is embedded in a space-varying external field.

\begin{figure}{}
\begin{center}
\resizebox{8.5cm}{!}{\includegraphics[angle=-90]{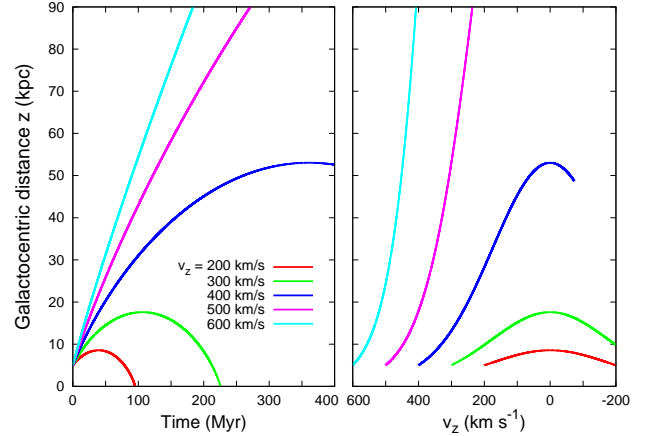}}
\makeatletter\def\@captype{figure}\makeatother
\caption{The radial orbits with different initial velocities starting from the position of $(x,~y,~z)=(0,~0,~5)\kpc$. {\bf Left panel:} Galatocentric distances versus orbital time. {\bf Right panel:} Galatocentric distances versus radial velocity.}\label{figorb}
\end{center}
\end{figure}

We choose a range of initial velocities $v_z=200\kms,~300\kms,~400\kms,~500\kms,~600\kms$, to move the GCs from the quasi-Newtonian regime to the Milgromian regime in the Galactic background field on different time scales.  We show the orbits in Fig. \ref{figorb}. We find that orbits with an initial velocity of $200\kms$ and $300\kms$ reach their apocentres at small radii, at about $8\kpc$ and $18\kpc$ respectively, and they cannot propagate to the outer part of the Galaxy. Thus these orbits are not of our interest. The dynamics of GCs moving on orbits with $v_z=~400\kms,~500\kms,~600\kms$ are studied in \S \ref{adiabatic}. We move the GCs on these orbits for $\approx 500\Myr$, which is long enough for the GCs to move to their apocentres or to the outer Galactic regime ($r\approx 90\kpc$) where the distant Galactic GCs are observed \citep{Baumgardt_etal2005,Bellazzini2007}. We point out that the apocentre for the orbits with $v_z=~400\kms$ is about $72\kpc$. We study the kinematics of the GCs at their apocentres or at Galatocentric distance of $90\kpc$, if their apocentres are even more distant. The orbital times from the initial position to the apocentre or to $z=90\kpc$ are: $360\Myr$ for $v_z=400\kms$, $270 \Myr$ for $v_z=500\kms$ and $180\Myr$ for $v_z=600\kms$. We note that the orbital time to transform the external field from the strong to the weak field is much longer than the re-virialisation time of $5\times \tdyn$, therefore the GCs are re-virialised gradually in the slow-evolving gravitional field. The phase transitions are adibatic.

For models $3$ and $5$, the starting points of the Galactic orbits are $(x,~y,~z)=(0,~0,~10)~\kpc$, and the models are moving on the same Galactic orbits as models $1,~2$ and $5$. The initial velocities are interpolated at the starting point from the above Galactic orbits (see Fig. \ref{figorb}).

\subsection{Adiabatically evolving systems}\label{adiabatic}
\subsubsection{Virial ratios}
\begin{figure}{}
\begin{center}
\resizebox{8.7cm}{!}{\includegraphics[angle=-90]{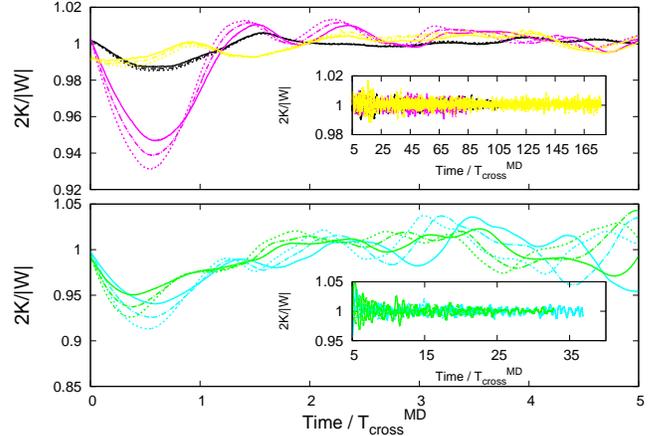}}
\makeatletter\def\@captype{figure}\makeatother
\caption{Upper panel: The virial ratio of models $1$ (black), $2$ (magenta) and $4$ (yellow) moving on radial orbits in the MW potential, with initial velocities of $400\kms$ (solid), $500\kms$ (dot-dashed) and $600\kms$ (dashed). Lower panel: The virial ratio of models $3$ (cyan) and $5$ (green) for comparison. The values of $\tcm$ can be found from Table \ref{ics}.}\label{vir_kick}
\end{center}
\end{figure}
The upper panel of Fig. \ref{vir_kick} shows the virial ratios of the GCs moving on the different orbits. We find that the systems are not violently re-virialised compared to the case of the rapid phase transition. The virial ratios only deviate by $7\%$ from $1$ at most within the first $5$ crossing times. For the compact models $1$ and $4$, the deviation is within $1\%$ in the first $5$ crossing times. Since the external fields are evolving, there is a noise of $1.5\%$ at $T \approx 18~\tcm$. Thereafter the fluctuations of virial ratios are within $1\%$ around $1$. Therefore the systems are only slightly out of virial equilibrium at the beginning of a Galactic orbit, where they are close to the Galactic centre and the gravitational field changes fastest. As we see from Fig. \ref{vir_kick}, from $5~\tcm$ onwards the systems are in equilibrium since the external field changes gradually. The differences of the time scales for systems reaching equilibrium states caused by different initial velocities can be ignored. Thus systems evolve adiabatically and the collapse process is much more moderate. The systems are very unlikely to be frozen in Newtonian dynamics when they move into the Milgromian regime, even when they are on an orbit with a high initial out-going velocity of as high as $600~\kms$.

The virial ratio of models $3$ and $5$ are shown in the lower panel of Fig. \ref{vir_kick}. The amplitudes of the deviations to virial equilibrium are about $7\%$, which is much smaller than that of the violent re-virialisation. The virial ratios oscillate around $1$ with a larger noise, since models $3$ and $5$ are originally further away from virial equilibrium at their initial Galactic position of $(x,~y,~z)=(0,~0,~10)\kpc$. And the time scale of their phase transition is very similar to the cases of models $1,~2$ and $4$: from $7~\tcm$ onwards the systems are in virial equilibrium. It confirms again that the freezing time scales in quasi-Newtonian dynamics are only a few $\tcm$, and the dynamics of the models quickly becomes Milgromian on their radial Galactic orbits.

\subsubsection{Lagrangian radii and mass profiles}
\begin{figure}{}
\begin{center}
\resizebox{8.7cm}{!}{\includegraphics[angle=-90]{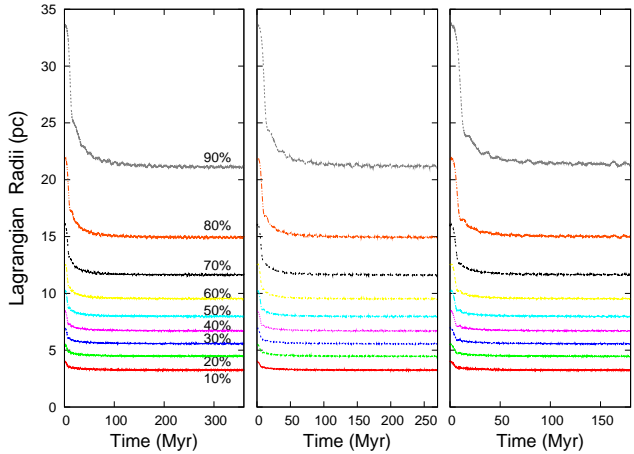}}
\makeatletter\def\@captype{figure}\makeatother
\caption{The evolution of Lagrangian radii of model $2$ on different Galactic orbits starting from $(x,~y,~z)=(0,~0,~5)\kpc$ and ending at their apocentres or at $r^{\rm MW}=90\kpc$ if the apocentres are even further away (see Fig. \ref{figorb}), with different initial velocities: $v_z=400\kms$ ({\bf left panel}), $500\kms$ ({\bf middle panel}), $v_z=600\kms$ ({\bf right panel}).}\label{rmass_kick}
\end{center}
\end{figure}

\begin{table*}
\begin{center}\vskip 0.00cm
\caption{Axis-ratios of the mass distribution of the GCs: The first column shows the ID of the models in Tab \ref{ics}, the second column shows the axis-ratios (see Eq. \ref{ixx}) of the GCs at $T~=~0$ near the Galactic plane, at the Galatocentric position $(0,~0,~5)\kpc$ for models $1,~2$ and $4$, and $(0,~0,~10)\kpc$ for models $3$ and $5$. The third to the fifth columns show the axis-ratios of the GCs in the outer regime of the Milky Way. The two parts of the table show the axis-ratios within $r_{50}$ (first three lines) and $r_{90}$ (last three lines). }
\begin{tabular}{llllllc}
\hline
Models& Inner MW & Apocentre& $r^{\rm MW}=90\kpc$ &$r^{\rm MW}=90\kpc$\\
  &  &$v_z=400\kms$ &$v_z= 500\kms$&$v_z=600\kms$\\
\hline
 & $(a:b:c)_{r_{50}}$ &  &  &\\
\hline
1 & $1: 0.98: 0.99$ &$1: 0.99: 0.99$  &$1: 0.99: 0.98$  &$1: 0.99: 0.99$\\
2 & $1: 0.98: 0.97$ &$1: 0.99: 0.99$  &$1: 1.00: 0.99$  &$1: 1.01: 1.01$\\
3 & $1: 0.96: 0.95$ &$1: 0.93: 0.92$  &$1: 0.97: 0.96$  &$1: 0.96: 0.97$\\
4 & $1: 0.97: 0.96$ &$1: 0.99: 0.99$  &$1: 1.00: 0.99$  &$1: 0.99: 0.99$\\
5 & $1: 0.95: 0.95$ &$1: 0.96: 0.96$  &$1: 0.98: 0.97$  &$1: 0.97: 0.96$\\
\hline
  & $(a:b:c)_{r_{90}}$ &  &  & \\
\hline
1 &$1: 0.98: 0.97$&$1: 0.99: 0.99$  &$1: 0.99: 0.99$  & $1: 0.99: 0.99$\\
2 &$1: 0.94: 0.94$&$1: 0.97: 0.98$  & $1: 0.99: 0.99$ & $1: 0.99: 0.99$\\
3 &$1: 0.88: 0.88$&$1: 0.88: 0.89$  & $1: 0.92: 0.92$ & $1: 0.91: 0.92$\\
4 &$1: 0.95: 0.95$&$1: 0.99: 0.99$  & $1: 0.99: 0.99$ & $1: 0.99: 0.99$\\
5 &$1: 0.89: 0.89$&$1: 0.92: 0.92$  & $1: 0.94: 0.94$ & $1: 0.93: 0.94$\\
\hline
\end{tabular}
\label{ell}
\end{center}
\end{table*}

We show the evolution of Lagrangian radii of model $2$ in Fig. \ref{rmass_kick}. The three panels correspond to different initial velocities of the Galactic orbits: $v_z=400\kms$ (left panel), $500\kms$ (middle panel) and $600\kms$ (right panel). The GC moves from the Galactic position of $(x,~y,~z)=(0,~0,~5)\kpc$ to the apocentre of the orbit (for the orbit of $v_z=400\kms$) or to $r^{\rm MW}=90\kpc$ where the most distant GC is observed. Compared to Fig. \ref{rmass_violent}, we find that the adiabatic evolution is significantly different to the violent evolution: the Lagrangian radii decrease fast during the first $\approx 20\tdyn^i$ (see Fig. \ref{tpt}), and then the Lagrangian radii are almost constant, with tiny oscillations. There are no significant oscillation of the radii within the first $20\tdyn^i$. This agrees with the evolution of virial ratios in Fig. \ref{vir_kick}. We also find that the final stable Lagangian radii of the same fraction of enclosed mass do not change with different Galactic orbits. I.e., models $2$ moving on different radial orbits have the same Lagangian radii when they are in the outer regime of the Galaxy. This implies that a GC in the outer regime of the MW will have a universal mass profile, which does not depend on the Galactic orbit. The evolution of Lagrangian radii of the other models are similar to that of model $2$.

\begin{figure}{}
\begin{center}
\resizebox{9.cm}{!}{\includegraphics{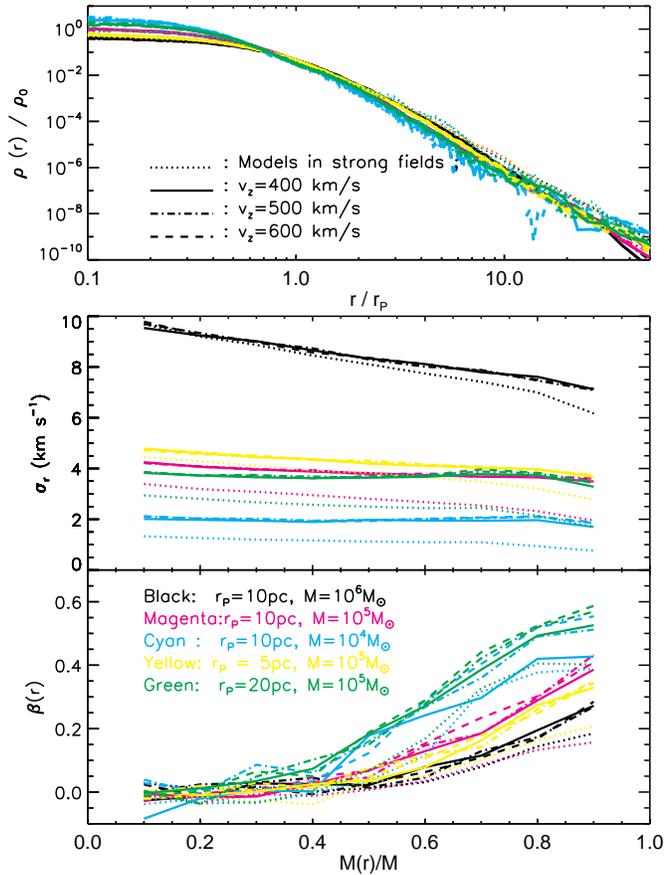}}
\makeatletter\def\@captype{figure}\makeatother
\caption{The spherically averaged density $\rho(r)$ ({\bf upper panel}) $\sigma_r(r)$ ({\bf middle panel}) and anisotropic profiles $\beta(r)$ ({\bf lower panel}) for models $1$ (black), $2$ (magenta) and $4$ (yellow) in a strong external field (dotted curves) and in weak fields. The systems move on radial polar orbits of the Galaxy, with different initial velicities, the line types showing the state of the systems at their apocentres: $v_z=400\kms$ (solid), $500\kms$ (dot-dashed) and $600\kms$ (dashed). The colours and line types are defined as in Fig. \ref{vir_kick}.}\label{prof_kick}
\end{center}
\end{figure}

We study the mass profiles of the GCs embedded in a strong field and in weak fields (at the apocentres of different orbits, or at $r^{\rm MW}=90\kpc$) in the upper panel of Fig. \ref{prof_kick}. We find that the densities of the GCs after adiabatic compression are very similar to the case of violent collapse: The densities in the core regime become larger while in the outer regime at $r>3r_{\rm P}$ they are smaller. For a same model, the final density profiles for different Galactic orbits are very similar.

We also study the axis-ratios of the $50\%$ and $90\%$ enclosed mass of the initial GCs at a Galatocentric position of $(0,~0,~5)\kpc$ and of the final products (GCs at large radii of the Galaxy). The axis-ratios are defined as $a:b:c=\sqrt{I_{xx}}:\sqrt{I_{yy}}:\sqrt{I_{zz}}$, where the $I_{xx},~I_{yy}$ and $I_{zz}$ are the diagonalised moments of inertia tensor directions \citep{Gerhard1983}:
\bey\label{ixx}
I_{xx} = \frac{1}{N_{\rm parts}}\sum_{\rm ip=1}^{N_{\rm parts}} \frac{(y_{\rm ip}^2+z_{\rm ip}^2)}{r_{\rm ip}^2},\nonumber \\
I_{yy} = \frac{1}{N_{\rm parts}}\sum_{\rm ip=1}^{N_{\rm parts}} \frac{(x_{\rm ip}^2+z_{\rm ip}^2)}{r_{\rm ip}^2},   \\
I_{zz} = \frac{1}{N_{\rm parts}}\sum_{\rm ip=1}^{N_{\rm parts}} \frac{(x_{\rm ip}^2+y_{\rm ip}^2)}{r_{\rm ip}^2}, \nonumber
\eey
where $r_{\rm ip}=\sqrt{x_{\rm ip}^2+y_{\rm ip}^2+z_{\rm ip}^2}$, and $N_{\rm parts}$ is the enclosed number of particles in the ellipsoids of $\left({x\over a}\right)^2+\left({y\over b}\right)^2+\left({z\over c}\right)^2=r^2_{50} ~({\rm or}~=~r_{90}^2)$, where $r_{50}$ and $r_{90}$ are the radii enclosing $50\%$ and $90\%$ of mass, respectively. The axes of the GCs are listed in Table \ref{ell}. When the models are at their starting points of the orbits, they are embedded in a strong external field, therefore the internal potentials of the GCs are prolate, especially for the outer parts \citep{Wu_etal2007,Wu_etal2008}. This is even more significant in models $3$ and $5$, since their internal structures are more diffuse than the other models.

We find that at their half mass radii (i.e., $r_{50}$), the models are slightly prolate since they are dominated by their internal fields at $r_{50}$. At the radii of $r_{90}$, the deviation from a spherical shape is larger, for model $2$ the axis-ratio is $1:~0.94:~0.94$. The GCs at the end of their out-going orbits are very close to be spherically symmetric after the compression process. There is only a deviation from a sphere of $1\%$ in both the inner and the outer parts of most of the models and for all radial orbits. For the moderately diffuse system, model $2$ on an orbit with $v_z=400\kms$, the axis-ratios at $r_{90}$ are slightly prolate, $1:0.97:0.98$. This is because model $2$ is more diffuse compared to the models $1$ and $4$, and at the radius of $r_{90}$ and the apocentre of the Galactic orbit, the external field can affect the outer parts. While model $2$ moves further away, say to $r^{MW}=90\kpc$, the external field is weaker than at the apocentre of the first Galactic orbit, and the model becomes spherically symmetric again.

Models $3$ and $5$ confirm this trend: they are intrinsically even more diffuse than model $2$, therefore they are even more prolate at both $r_{50}$ and $r_{90}$. However, the deviations from spherical symmetry are next to negligible for all models.

\subsubsection{Velocity dispersion and anisotropy}\label{aniso-kick}
The radial velocity dispersion profiles are studied in the middle panel of Fig. \ref{prof_kick}. The shapes of the final radial velocity dispersion profiles are very similar to those of the violent re-virialisation. However, there are small differences between the profiles for GCs collapsing violently and adiabatically, especially for the diffuse GCs, model $2$, and also the most diffuse models $3$ and $5$. Compared to the middle panel of Fig. \ref{prof_violent}, models $2$, $3$ and $5$ have increasing and then decreasing profiles in the intermediate regime between $r_{50}$ (where $\frac{M(r)}{M}=0.5$) and $r_{90}$ (where $\frac{M(r)}{M}=0.9$) for the violent collapse cases, while the $\sigma_r(r)$ profiles are almost flat for the case of adiabatic collapse. In the other two models the $\sigma_r(r)$ profiles are almost the same for adiabatic and violent collapses. Moreover, we find that the $\sigma_r(r)$ profiles of the final products are independent of their Galactic orbits. Different Galactic orbits lead to a similar $\sigma_r(r)$ profile when the GCs are in the outer regime of the Galaxy.

There are more differences in the anisotropy profiles when comparing with the case of violent re-virialisation: the GCs become much more radially anisotropic if they collapse violently. This is very clear from comparing the lower panels of Figures \ref{prof_violent} and \ref{prof_kick}. For model $2$, $\beta(r_{90})=0.8$ for violent compression and $\beta(r_{90})$ is only $0.4$ for adiabatic compression. For the most diffuse models $3$ and $5$ this is even more significant: values of $\beta(r_{90})$ are $0.85-0.90$ for violent re-virialisation while the values of $\beta(r_{90})$ are only $0.45-0.6$ for adiabatic compression. The more compact models (model $1$ and $4$) have the same trend although not as significant as model $2$, since among the three models, model $2$ itself is in deeper MD.

Moreover, the final $\beta(r)$ profiles for different GCs are independent of the Galactic orbits. There is only one exception, model $3$, for which the difference in anisotropy, $\delta \beta(r_{90})$ is up to $0.2$. The large $\delta \beta(r_{90})$ comes from there being fewer particles for model $3$ therefore the particle noise is larger.

Thus the anisotropy profiles are related to the internal structures of the GCs. The faster the systems move from the inner to the outer Galaxy, the more radially anisotropic the systems are. However, both the $\sigma_r(r)$ profiles and the $\beta(r)$ profiles are not related to the orbital history. I.e., the internal kinematics of a GC, which is moving on an eccentric Galactic orbit and is currently in the Milgromian regime, only depends on the internal structure of the GC rather than on the details of the orbital history. This is good news, and it makes the predictions from Milgromian dynamics less complex.

\subsubsection{Phase space distribution with adiabatic re-virialisation}

\begin{figure}{}
\begin{center}
\resizebox{9.cm}{!}{\includegraphics{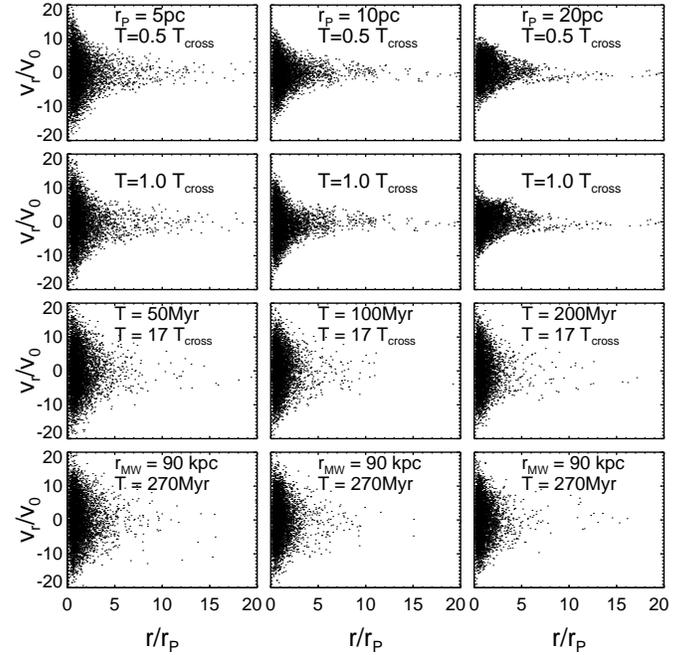}}
\makeatletter\def\@captype{figure}\makeatother
\caption{The phase space distribution at different snapshots of adiabatic re-virialisation for models $4$ (left panels), $2$ (middle panels) and $5$ (right panels) moving on a Galactic orbit with initial position of $5~\kpc$ and initial velocity of $500~\kms$ (purple curves in Fig. \ref{figorb}, for model $5$ the Galactic orbit is the same one but the starting point is $10~\kpc$ and the initial velocity is interpolated from the purple curve in the right panel of Fig. \ref{figorb}). The radial radii are scaled by the Plummer radius $r_{\rm P}$, and the radial velocity of particles $v_r$ are scaled by $v_0=\sqrt{GM/r_{\rm P}}$.}\label{ps_kick500}
\end{center}
\end{figure}

The phase space distributions of systems with adiabatic
re-virialisation are presented in Fig. \ref{ps_kick500}, which can
be compared with the case of violent re-virialisation. One Galactic
orbit is selected from Fig. \ref{figorb}, the purple curves, i.e.
the initial Galactic position for the cluster is at $r_{\rm
MW}~=~5\kpc$ and the initial velocity is $500~\kms$. We show in
Fig. \ref{ps_kick500} the models $4$ (left panels), $2$ (middle
panels) and $5$ (right panels) at time snapshots of $0.5~\tcm$,
$1.0~\tcm$, $17~\tcm$ and the apocentre of the Galactic orbit (or the furthest
position that Milky Way star clusters are observed, say, $90~\kpc$
away to the Galactic centre). At the early stage of
re-virialisation, the systems appear to be well phase mixed. Only
for the deep-Milgromian system, model $5$, there are structures with
negative radial velocity in the outer regimes of $r~>~ 5~ r_{\rm
P}$, which indicates that a large amount of particles are falling
into the centre of the star cluster. The reason is that, on a
realistic Galactic orbit, the self-potential of the star cluster
does not change to a sufficiently significant extent during one
crossing time of the cluster. Only when the star cluster has a very diffuse mass distribution and the self-potential is dominated by deep-Milgromian dynamics, a small amount of change of the background external field is abe to yield a visible effect in the phase space distribution. The systems are already well phase mixed at $17~\tcm$, and the phase space distributions are very close to that of the final time snapshot, namely at the furthest Galactic position where the star cluster can reach (the bottom panels of Fig. \ref{ps_kick500}).

\begin{figure}{}
\begin{center}
\resizebox{9.cm}{!}{\includegraphics{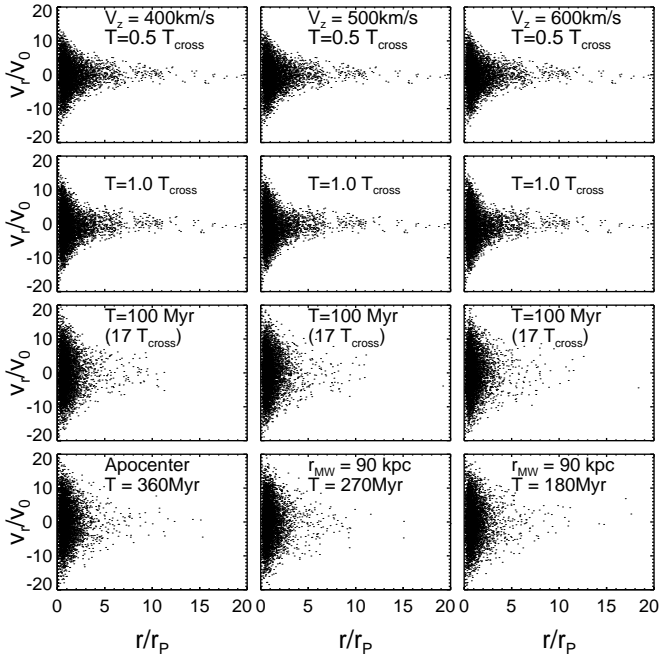}}
\makeatletter\def\@captype{figure}\makeatother
\caption{The phase space distribution at different snapshots of adiabatic re-virialisation for model $2$ moving on different Galactic orbit with initial position of $5~\kpc$ and initial velocity of $400~\kms$ (left panels), $500~\kms$ (middle panels) and $600~\kms$. The radial radii are scaled by the Plummer radius $r_{\rm P}$, and the radial velocity of particles $v_r$ are scaled by $v_0=\sqrt{GM/r_{\rm P}}$.}\label{ps_kick1e5}
\end{center}
\end{figure}

There is a wide range for the initial conditions of Galactic orbits
for the GCs. Thus it is important to study the phase space
distribution for GCs on different Galactic orbits. Model $2$ is
presented on different Galactic orbits with different initial
velocities in Fig. \ref{ps_kick1e5}: $400~\kms$ (left panels),
$500~\kms$ (middle pannels) and $600~\kms$ (right panels). On an
orbit with higher initial velocity, it is clear there are in-falling
particles at an early stage of $T~=~1.0~\tcm$. The systems are already
well phase mixed at $17~\tcm$, although the GCs are still moving on
the orbits. It agrees with the results of GCs with different sizes
moving on a same orbit in Fig. \ref{ps_kick500}.

We therefore conclude that, for the systems re-virialised
adiabatically, the phase mixing is more efficient than that in the
violent re-virialisation process. The self-potentials for GCs are
mildly deepened in the former case, because the
systems are very close to being in equilibrium on the orbits.

\subsubsection{Stability of the star clusters after adiabatic re-virialisation}

The study of the anisotropy of the systems reveals that GCs are
less radially anisotropic through adiabatic re-virialisation
compared to that realized through violent re-virialisation. It is
therefore reasonable to expect $\xi$ parameters (Eq. \ref{xi}) to be smaller for
the systems adiabatically re-virialised. The values of $\xi$ for GCs
re-virialised on different Galactic orbits are listed in Table
\ref{stab-kick}. For all the models $\xi<\xi_c$ \citep[The critical
$\xi_c$ values in Milgromian dynamics is emperically calculated in
][]{Nipoti_etal2011}. Apparently, all the GC systems are stable
after adiabatic re-virialisation. It can also be found from Table
\ref{stab-kick} that for the more compact GCs (Models $1$, $2$ and
$4$), the $\xi$ values are very similar to each other on different
Galactic orbits. For the more diffuse GCs (Models $3$ and $5$) which
re-virialised into deep-Milgromian gravity, the difference of $\xi$
values are more significant on different Galactic orbits. For
systems with larger initial velocity, the $\xi$ values are larger,
which is owing to the presence of more radial orbits in the systems which
re-virialised more rapidly. These results agree well with the
analysis of anisotropy in \S \ref{aniso-kick}.

The values of $\xi$ for GCs after adiabatic re-virialisation are
smaller than those in the case of violent re-virialisation,
suggesting that adiabatically re-virialised GCs are even more
stable.

\begin{table*}
\begin{center}\vskip 0.00cm
\caption{The stability parameter $\xi$ for the GC models adiabatically re-virialised from the Galatocentric positions ($(0,~0,~5)\kpc$ for models $1,~2$ and $4$, and $(0,~0,~10)\kpc$ for models $3$ and $5$) to the outer regime of the Milky Way. The second to the forth columns show the values of $\xi$ of the clusters moving on different Galactic orbits.}
\begin{tabular}{llllllc}
\hline
Models& Apocentre& $r^{\rm MW}=90\kpc$ &$r^{\rm MW}=90\kpc$\\
      &$v_z=400\kms$ &$v_z= 500\kms$&$v_z=600\kms$\\
\hline
1 & 1.09 & 1.09 & 1.09 \\
2 & 1.16 & 1.17 & 1.19 \\
3 & 1.24 & 1.32 & 1.35 \\
4 & 1.12 & 1.12 & 1.13 \\
5 & 1.29 & 1.34 & 1.36 \\
\hline
\end{tabular}
\label{stab-kick}
\end{center}
\end{table*}

\section{Conclusions and discussions}\label{conc}

We study the violent and adiabatic re-virialisation of stellar systems from Newtonian to Milgromian gravity. The time scale of the re-virialisation is only a few crossing times in MD, for both the violent and the adiabatic process. Therefore it would be unlikely to find systems like diffuse GCs or ultra faint dwarf galaxies that are in a weak background field regime to be frozen in Newtonian dynamics. There are GCs like Pal 4 \citep{Frank_etal2012}, Pal 14 \citep{Jordi_etal2009} and NGC 2419 \citep{Baumgardt_etal2009,Ibata_etal2011b,Ibata_etal2011a}, which at first sight appear to be Newtonian diffuse systems in the outer Galactic regime. We conclude that this behaviour cannot be explained by these systems being frozen in Newtonian gravity. The adiabatic simulations show that the orbital time is much longer than the dynamical times $\tdyn^i$ of the systems staying out of equilibrium. A more recent study on NGC 2419 \citep{Sanders2012a,Sanders2012b} shows that NGC 2419 is not a problem for MD. \citet{Sanders2012a,Sanders2012b} used polytropic models to model NGC 2419, which fit well the observations of the surface brightness and kinemaics. However \citet{Ibata_etal2011a} claimed that polytrope models are less likely in Milgromian dynamics by a factor of 5000 than a Newtonian Michie model as used in \citet{Ibata_etal2011b}. Therefore currently NGC2419, Pal 4 and Pal 14 cannot be explained by the potential being frozen in its Newtonian form. 

We also study the Lagrangian radii, mass profiles and kinematics of the final products for the above two re-virialisation process. We find that the mass profiles and radial velocity dispersion profiles are very similar for the systems collapsing the two different ways. Moreover, different Galactic orbits for a GC lead to the same Lagrangian radii, $\rho(r)$ and $\sigma_r(r)$ profiles. Therefore for a system moving from the inner regime of the Galaxy to the outer part, the Lagrangian radii, mass profile and $\sigma_r(r)$ are independent of the history of the GC's orbit.

In the potential of the Galaxy, we revirialise the spherically symmetric Newtonian systems in the inner Galactic field using Milgromian dynamics. The velocity dispersion becomes radially anisotropic after the re-virialisation due to the external field effect: since the SEP is broken, the internal potential of a GC is asymmetric, therefore the orbits of the particles in such a GC are distorted and elongated. We then move the GCs from the inner to the outer regime of the Galaxy. We find that in those GCs, the velocity dispersion profiles should evolve to even more radially anisotropic profiles. Note that the $\beta(r)$ profiles are determined by the internal structure of the GCs rather than by the details of the Galactic orbits: the more diffuse the GCs are, the more radially anisotropic their velocity dispersions are after the re-virialisation. The GCs on an orbit moving faster or slower from the Newtonian regime to the Milgromian regime have similar radially anisotropic velocity dispersion profiles. The re-virialisation is a new mechanism to produce such profiles compared to Newtonian dynamics. In contrast, Newtonian N-body models of star clusters generate isotropic or mildly anisotropic velocity dispersion profiles.

Observations of the distant GCs NGC 2419 by
\citet{Baumgardt_etal2009} and
\citet{Ibata_etal2011b,Ibata_etal2011a} show that the line-of-sight
velocity dispersion profile of NGC 2419 has large values up to
$7~\kms$ in the centre and small values of $1-2\kms$ in the outer
regime where $r>3r_h$ \citep[see Fig. 8 of ][]{Ibata_etal2011b}.
The studies on the dynamics of NGC 2419 require a highly radially
anisotropic velocity dispersion to fit the observed data in both
Newtonian or Milgromian gravities
\citep{Ibata_etal2011b,Ibata_etal2011a}. In Newtonian dynamics,
one mechanism to generate such radial anisotropy is partial
relaxation, i.e. a violent relaxation process that is inefficient in
the outer parts of the system \citep{Lynden-Bell1967,
Bertin_Trenti2003}. A self-consistent family of radially
anisotropic models of partially relaxed systems have been proposed
in \citet{Bertin_Trenti2003}. Later, \citet{Trenti_etal2005}
compared such a family of models and the products of collisionless
collapse. It was confirmed that these models are unstable while the
stability parameter $\xi > 1.7\pm 0.25$ in Newtonian dynamics, and
that the strongly radial models will evolve into triaxial systems.
Recently, \citet{Zocchi_etal2012} investigated these models for a
sample of GCs including NGC 2419. The stability parameter is found
to be $\xi=1.77$ for the best fit model of NGC 2419, so NGC 2419 is
on the boundary of being stable in Newtonian dynamics.

The phase transition provides another mechanism for generating the
anisotropy in Milgromian dynamics, in addition to the partial
relaxation which has not yet been studied in Milgromian dynamics.

\begin{figure}{}
\begin{center}
\resizebox{9.cm}{!}{\includegraphics{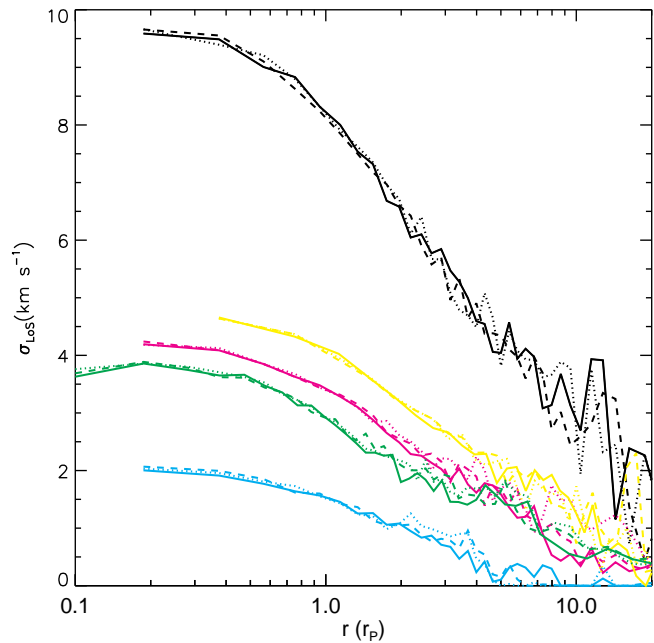}}
\makeatletter\def\@captype{figure}\makeatother
\caption{The final line-of-sight velocity dispersion profiles as a function of radius for the models after moving from the inner Galaxy to the outer Milgromian regime. The colours and line types are defined as in Fig. \ref{vir_kick}.}\label{sig_los_kick}
\end{center}
\end{figure}

Here we also show the final line-of-sight velocity dispersion profiles, $\sigma_{LOS}(R)$, of our GC models after moving from the inner Galaxy to the outer Galactic regime (Fig. \ref{sig_los_kick}). Since our GC models are radially anisotropic, especially for models $2, 3$ and $5$, the shapes of their $\sigma_{LOS}(R)$ should be similar to that of NGC 2419. Indeed we find that the $\sigma_{LOS}(R)$ have large values in the centres of the GCs and sharply fall from $1r_P$ to $10r_P$. The $\sigma_{LOS}(R)$ profiles are very similar for the same model moving on different Galactic orbits (see Fig. \ref{sig_los_kick}). It confirms that the observational results of line-of-sight kinematics of a GC are not related to the orbital history.

In a follow-up project we will study other possible mechanisms for
generating radially anisotropic velocity dispersion profiles and the
corresponding $\sigma_{LOS}(R)$ profiles in both MD and Newtonian
gravities. For instance, it is interesting to consider gas
expulsion after birth and mass loss from evolving mass stars in the early stage of the GCs \citep[][for initial conditions of GCs]{Marks_Kroupa2012} and
make a comparison between the different radial anisotropic
behaviours generated from Newtonian and Milgromian gravities.

In summary, Milgromian dynamics predicts that all of the GCs moving on eccentric orbits of the Galaxy should be radially anisotropic when they are in the Milgromian regime. Any isotropic or tangentially anisotropic profiles for GCs in the outer regime of any radial Galactic orbit will be problematic for MD, since these systems experience a collapse during the phase transition. Therefore for any out-going systems, any observations on isotropy or tangential anisotropy will be a severe challenge to MD. MD is falsifiable by the kimematics of the outer GCs.

\section{Acknowledgments}
Xufen Wu gratefully acknowledges support through the Alexander von Humboldt Foundation. We thank Ortwin Gerhard and Flavio de Lorenzi for sharing the code for generating anisotropic models by using circularity functions and Lucy's method. We also thank the Bologna group, Nipoti, Ciotti and Londrillo for sharing their N-body code NMODY.

\appendix
\section{Stability test of unperturbed Newtonian Initial Conditions (ICs)}\label{stability}
\begin{figure}{}
\begin{center}
\resizebox{8.7cm}{!}{\includegraphics[angle=-90]{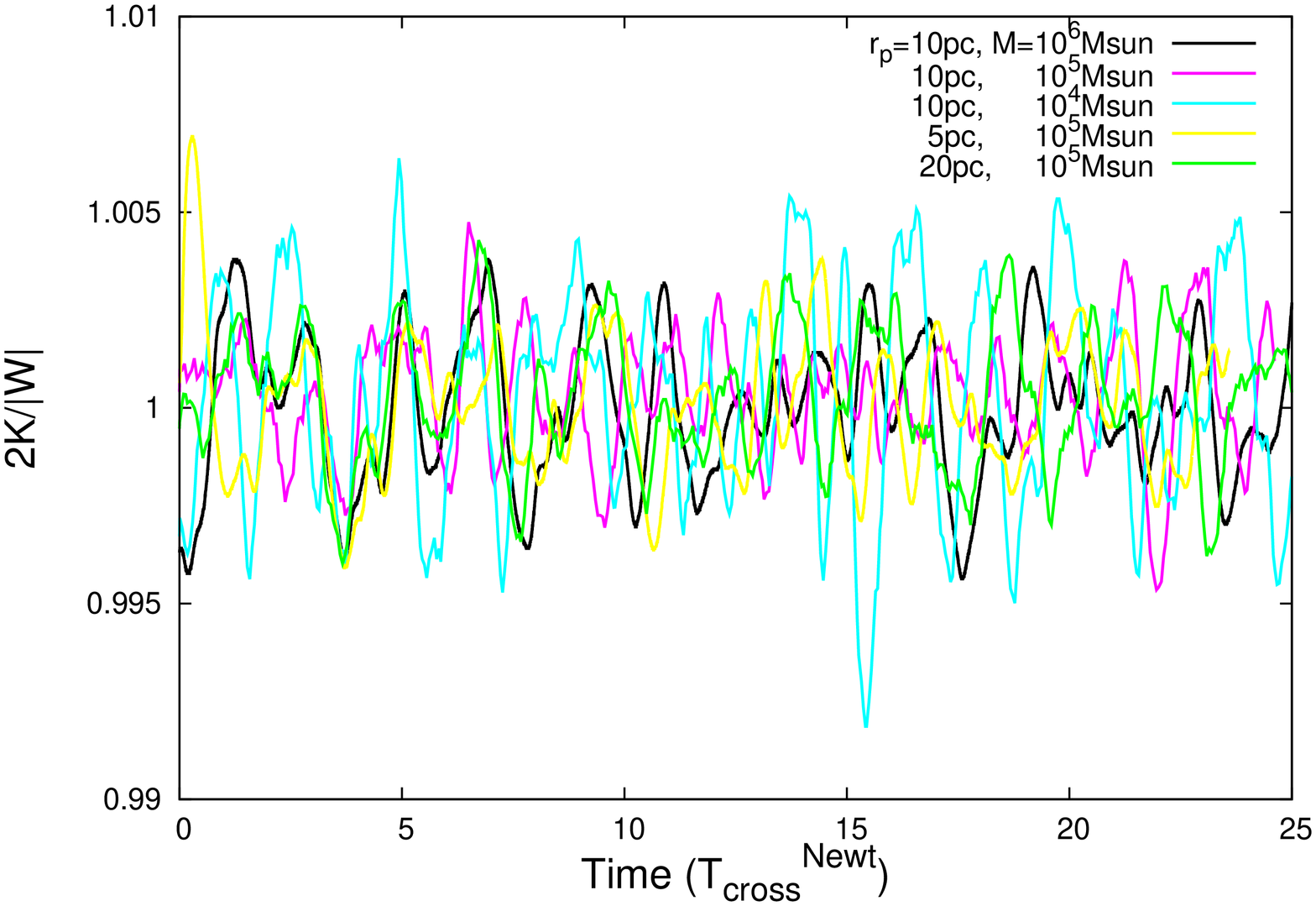}}
\makeatletter\def\@captype{figure}\makeatother
\caption{The virial ratio during the free evolution of the Newtonian models.}\label{stabtest1}
\end{center}
\end{figure}

We freely evolve the Newtonian ICs constructed in \S \ref{egc} in their self-potentials. Fig. \ref{stabtest1} shows the scalar virial ratios $\vir$ of the models evolving for about $25~\tcn$. We find that the virial ratios are always around $1$ with small perturbations of amplitude $\delta (\vir) < 1\%$. The models are therefore in virial equilibrium.

\begin{figure}{}
\begin{center}
\resizebox{9.cm}{!}{\includegraphics{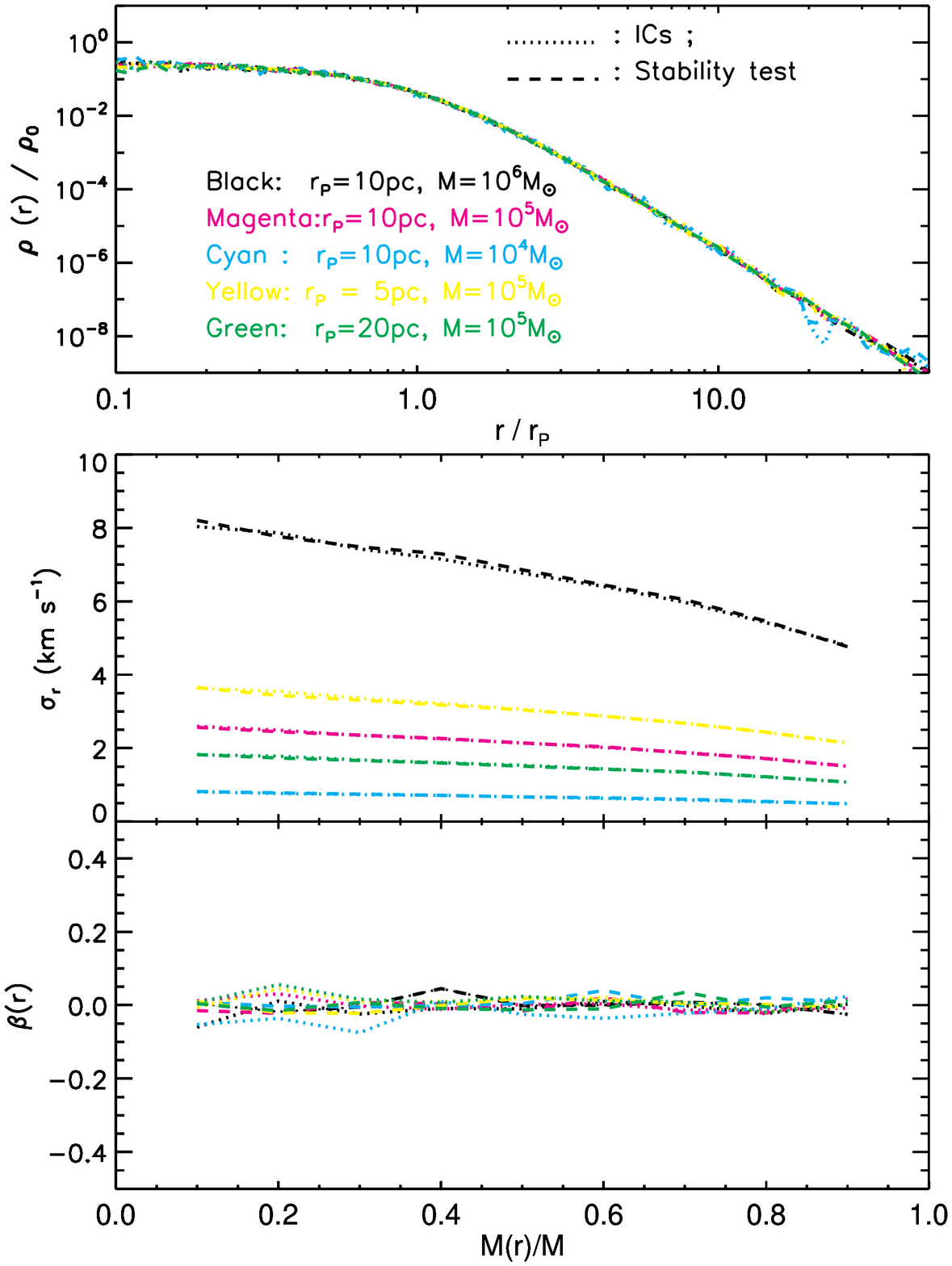}}
\makeatletter\def\@captype{figure}\makeatother
\caption{The spherically averaged density (upper panel), radial velocity dispersion (middle panel) and anisotropic profiles (lower panel) for models in Table \ref{ics}. The  dotted and dashed lines are for ICs and the models which are evolved for about $25~\tcn$ in Newtonian gravity.}\label{stabtest2}
\end{center}
\end{figure}

We further show the spherically averaged density profiles of the models before (dotted lines) and after (dashed lines) the phase mixing in the upper panel of Fig. \ref{stabtest2}, as well as their radial velocity dispersion (middle panel) and anisotropy profiles (lower panel, Eq. \ref{beta}). We find that their density, radial velocity dispersion and anisotropy profiles do not evolve with time. The models are still isotropic models after $25~\tcn$ and their spatial and velocity distributions do not change. Thus we conclude that the phase space distribution functions of the models do not evolve during the free evolution, and the models are stable. This is expected for collisionless systems.

\bibliographystyle{mn2e}
\bibliography{ref}
\end{document}